%% file: main.tex
\documentclass[11pt]{article}
\usepackage[utf8]{inputenc}
\usepackage{url,geometry}
\usepackage[numbers]{natbib}

\usepackage{adjustbox}
\usepackage[figuresright]{rotating}
\usepackage{amsmath}
\usepackage{csquotes}
\usepackage{appendix}
\usepackage{booktabs,array}
\usepackage{multirow}
\usepackage{type1cm}
\usepackage{url}
\usepackage[group-separator={,}]{siunitx}
\usepackage[table,xcdraw]{xcolor}
\usepackage{lscape}
\usepackage{graphicx}
\usepackage{setspace}
 \usepackage[normalem]{ulem}
 \useunder{\uline}{\ul}{}
 \usepackage{colortbl}
\usepackage{tabularray}
\usepackage{authblk}

\usepackage[final]{changes}

\title{Voting by mail: a Markov chain model for managing the security risks of absentee voting}
\author[1]{Carmen A. Haseltine \thanks{carmen.haseltine@morgan.edu}}
\author[2]{Laura A.~Albert\thanks{laura@engr.wisc.edu}}
\affil[1]{Department of Electrical \& Computer Engineering, Morgan State University}
\affil[2]{Department of Industrial \& Systems Engineering, University of Wisconsin-Madison}

\date{To appear in \emph{Decision Analysis} \\ March 31, 2026}

\begin{document}
\maketitle


\begin{abstract}
The scrutiny surrounding vote-by-mail (VBM) in the United States has increased in recent years, \added{highlighting the need for a rigorous quantitative framework to evaluate the resilience of the absentee voting infrastructure.} This paper addresses these issues by introducing a dynamic mathematical modeling framework for performing a risk assessment of VBM processes. We introduce a discrete-time Markov chain (DTMC) to model the VBM process and assess election performance and risk with a novel layered network approach that considers the interplay between VBM processes, malicious and non-malicious threats, and security mitigations. The time-inhomogeneous DTMC framework captures dynamic risks and evaluates performance over time. The DTMC model accounts for a spectrum of outcomes, from unintended voter errors to sophisticated, targeted attacks, representing a significant advancement in the risk assessment of VBM planning and protection. A case study based on real-world data from Milwaukee County, Wisconsin, is used to evaluate the DTMC model. The analysis includes \added{hypothetical worst-case attack scenarios to stress-test VBM processes and to assess the efficacy of security measures and} the impact of different attack timings. The analysis suggests that ballot drop boxes and automatic ballot notification systems are crucial for \added{reducing the attack surface to ensure} secure and reliable operations.
\end{abstract}


\maketitle

\section{Introduction}\label{sec:intro}
The legitimacy of any political system hinges on the accuracy and fairness of its voting processes.
In the United States (US), there are generally two methods by which the public can cast a ballot. Voters can travel to their designated polling location where they receive and cast a ballot on an election day, referred to as in-person voting. Voters can also vote absentee, and many of those who vote absentee choose to accept and cast a ballot through the United States Postal Service (USPS), the vote-by-mail (VBM) system. Historically, in-person voting has been the primary method of voting in the US. However, VBM has experienced significant growth in recent years, accounting for 20-percent of votes cast in the 2016 General Election in the US and 46-percent of votes cast in the 2020 General Election that occurred during the COVID-19 pandemic \cite{Stanford_VBMsecure}.

The popularity of VBM among voters has highlighted the need for an efficient, reliable, and secure plan for this voting modality.
The increased volume of VBM ballots in the 2020 General Election revealed issues such as mailing delays, security concerns, and accessibility to voting infrastructure. Since 2020, Colorado, Oregon, and six other states have adopted ``all-mail'' voting, which involves pro-actively sending mail-in ballots to all registered voters \cite{VBM2022}. Additionally, many US states have proposed or are considering legislation that has implications for VBM operations and security \cite{NCSL}. \added{While election processes are believed to be secure in the United States \cite{krebs2020,EAC2025},} election planning requires an understanding of the security needs crucial for VBM processes and, by extension, core election infrastructure. 

The VBM process involves a complex interplay of time-sensitive tasks among election offices, voters, and the USPS. These tasks comprise the election procedures and policies managed at the state and local levels. As a result, VBM processes can vary depending on the location.
\added{While there is extensive preparation prior to every election, it is not practical to physically stress-test the entire VBM infrastructure,} which drives the need for enhanced security planning tools and risk analysis models. Additionally, the VBM process occurs over several months, making it vulnerable to temporal issues occurring at specific times, such as USPS mail processing delays causing ballots to arrive after an election day when they are no longer counted (in many US states), as well as in-person absentee voting that is only available on specific days leading up to an election. We integrate these risks into a dynamic mathematical model of the VBM system capable of evaluating different threats and measures to mitigate them.

This paper presents a methodology to support VBM risk analysis by introducing a novel discrete-time Markov chain (DTMC) for VBM to capture dynamic election performance \added{and stress-test VBM processes}. The modeling approach uses the attack tree models of malicious (intentional) and non-malicious (unintentional) risks and security mitigations. 
Attack tree models have been used to characterize risks in prior studies on voting by mail security \cite{EAC_2009,scala2022evaluating}.
A novel aspect of the DTMC framework is that it models the linkages between attack trees, security mitigations, and VBM processes to characterize and predict dynamic VBM performance under various threat scenarios and operating conditions.

\added{The framework captures a predefined set of threats and VBM process rules to examine hypothetical worst-case scenarios to test the ability of security mitigations to withstand these targeted attacks independent of the historical rarity of such events. The case study analyses seek to stress-test the current security landscape rather than provide a forecast of daily operations.}
The DTMC framework highlights how different types, intentions, and timings of attacks impact overall VBM performance and security. We acknowledge that successful attacks could \added{potentially} change the outcome of an election.
Using real-world data from Milwaukee County in the 2020 General Election, we evaluate VBM performance under different conditions and policy scenarios. 

We focus on the number of ballots affected by election risks, not on the likelihood of the outcome of an election changing, since any inaccuracies weaken the legitimacy of a political system, regardless of whether these inaccuracies change election outcomes.
\added{From a decision analysis perspective, minimizing the expected number of ballots affected by risk serves as a means objective. While the primary fundamental objective of an adversary is typically viewed as preserving the election outcome (i.e., minimizing the probability that the true winner is not certified), other fundamental objectives for an adversary may be to undermine public trust in the democratic process or delegitimize the results, regardless of the final count \cite{stewart2022trust,cisa2024_insider}. Under the trust erosion objective, for example,  the adversary does not need to mathematically flip the election outcome; they only need to generate enough affected ballots (e.g., lost, stolen, or successfully challenged votes) to fuel a narrative of systemic failure. Therefore, our focus on the number of affected ballots supports a robust defense strategy, since it simultaneously hardens the system against outcome-reversal attacks, which require affected volume to exceed the margin of victory, and trust-erosion attacks, which require affected volume to exceed the public's tolerance for error.
}

Contributions of this paper include the following. 
\begin{enumerate}
    \item We formulate a new DTMC using a layered network approach that evaluates election performance and risk over time. \added{The model is derived from an existing VBM process map \cite{ScalaProcessMap}, which aggregates granular operational steps into functional states to allow for a quantification of ballot flow and risk measurement across the voting timeline.}
    \item We construct a new dataset of VBM mitigations and their attributes to link each mitigation to attacks and the VBM process.
    \item \added{We introduce a methodology for stress-testing VBM processes. Given that historical security incidents such as confirmed fraud cases are statistically rare, empirical threat probabilities are difficult to estimate. Therefore, we consider hypothetical worst-case scenarios to evaluate the system's resilience against potential disruptions.}
    \item We perform a detailed analysis of the DTMC model using a real-world case study based on Milwaukee County, Wisconsin that considers multiple, \added{hypothetical} threat scenarios to \added{stress-test the model and} shed light on how mitigations impact performance of the VBM process.
\end{enumerate}

The structure of this paper is as follows. Section \ref{sec:litreview} summarizes the existing research on election security modeling. Section \ref{sec:layers} summarizes the components of VBM, and Section \ref{sec:mathmodel} introduces the DTMC and performance measures. Section \ref{sec:casestudy} details the case study of Milwaukee County, Wisconsin. 
Section \ref{sec:results} summarizes the case study findings, and the paper concludes in Section \ref{sec:conclusion} with policy implications.

\section{Literature review }\label{sec:litreview}

The vast majority of election security modeling research focuses on in-person voting. A stream of papers in this area assesses the impact of operational decisions and resource allocation on election performance using methodologies such as discrete event models and model optimization  \cite{stewart2015waiting,stewart_survey_2020}.
However, these approaches are not adaptable to address VBM, which requires its own analytical framework due to its unique operations and vulnerabilities that differ from in-person voting. An exception is  \citeauthor{adamDropbox} \cite{adamDropbox} who formulate an integer programming model to determine the locations of drop boxes used in VBM. However, \citeauthor{adamDropbox} \cite{adamDropbox} narrowly focus on the location of drop boxes and the collection of ballots from drop boxes. In contrast, our paper focuses on the entire VBM process, including operations leading up to an election.

\added{ Crimmins et al.\cite{crimmins2025improving} use robust optimization to develop efficient logic and accuracy testing to ensure that voting machines accurately count votes. }

Research on the allocation of voting machines to polling locations for in-person voting has been extensively explored through various methodologies. One notable example is a study by \citeauthor{yang2009all} \cite{yang2009all} who uses discrete event simulation to model election day processes, combining queuing theory and optimization to address resource concerns. 
Similarly, \citeauthor{li2013could} \cite{li2013could} employ simulation optimization to assess the impact of voting machine distribution on voter queue times. The significance of these studies extends to a tailored risk assessment for VBM. \citet{adamCovidVoting} apply discrete event simulation to understand the effects of safety measures implemented during the COVID-19 pandemic on wait times and other election performance indicators. In addition, \citet{queueingModelMIT} focus on optimizing the polling locations to reduce voter waiting times. These approaches underscore the importance of modeling and optimization in improving election processes, informing the development of risk assessment methodologies for election infrastructure.

Many ballots by mail are tallied by scanning ballots by voting equipment. To ensure that votes are accurately counted, Crimmins et al. \cite{crimmins2025improving} use robust optimization to develop efficient logic and accuracy testing. Specific to the VBM system, \citet{ScalaProcessMap}  develop a detailed model of the VBM process, which meticulously outlines its various components and the physical journey of ballots. The Cybersecurity and Infrastructure Security Agency (CISA) conducted a comprehensive assessment of mail-voting security for the 2020 General Election \cite{cisa2020_VBM_risk}, contributing to efforts to identify and mitigate VBM-related risks during the pandemic. 

A growing body of literature characterizes threats to the voting process using attack trees. Attack trees help identify and prioritize potential vulnerabilities in a system, making them an essential tool in cybersecurity planning and risk mitigation. Attack trees are fundamental in cybersecurity for visualizing vulnerabilities and potential attacker paths \cite{attacktrees}.
An attack tree starts with the primary goal of the attacker (the root) and branches out into different methods or steps that the attacker might employ (the leaves). This branching structure often incorporates logic gates like AND and OR to demonstrate how different actions might combine to lead to the attack goal. Terminal attack nodes, are the leaf nodes of an attack tree that initiate the attack on the system. 
The \citet{EAC_2009} present a risk analysis report for election systems and models security threats to the VBM using attack trees that capture all possible vulnerabilities in the VBM system.
\citet{scala2022evaluating} expand on the VBM threat trees and adds to tree logic enumerating all terminal attack nodes to account for the new mechanisms available, such as drop boxes and in-person absentee voting. \citet{iise_cah}  build upon this body of literature to identify mitigation strategies and policies that protect the VBM process, accounting for linkages to attack trees relevant to VBM processes.
Our DTMC framework utilizes these attack trees and mitigations.

Attack and fault trees have been used for the risk analysis of in-person voting processes. \citeauthor{UA_threatTrees1} \cite{UA_threatTrees1} illustrate how attack trees have been deployed to analyze risks, including malicious cyber-attacks targeting voting machines. These methodologies consider both equipment failures and cybersecurity vulnerabilities, offering a broad view of potential threats. In a similar vein, \citeauthor{FaultTree} \cite{FaultTree} use fault trees to evaluate the in-person election process, with a particular emphasis on ballot counting and vulnerability detection. However, these papers do not differentiate between malicious and non-malicious attacks, solely focus on in-person voting, and represent risks as static by excluding temporal analysis. 

Our study aims to fill gaps in knowledge by introducing a time-inhomogeneous DTMC model to support VBM risk analysis over an election cycle. The DTMC model differentiates between malicious and non-malicious threats, considers risk in a dynamic system, and evaluates the effectiveness of mitigation strategies. This DTMC modeling approach employs layered networks to model interdependent components within the VBM process. The DTMC model is uniquely tailored to address the inherent complexities of VBM, efficiently accommodating simplified process components, diverse attack scenarios, and corresponding mitigation strategies.
Furthermore, the integration of VBM procedures within a cyber-physical systems (CPS) framework highlights the processes and threats that span cyber and physical components as well as the interdependence of the overall system. This is consistent with the observations of \citeauthor{Rinaldi} \cite{Rinaldi}, who examine the interconnected nature of modern critical infrastructure. Voting systems are integral to social and political structures and therefore require a holistic security approach to maintain their integrity \cite{Saltman1988VotingIntegrity}. Our paper aligns with these perspectives to provide a comprehensive and dynamic approach to safeguarding the voting process.

\section{Markov Chain Framework}\label{sec:layers}
We introduce a DTMC framework to evaluate VBM performance and assess its associated risks which captures the dynamic nature of the VBM process that unfolds over several months. We adopt a layered approach in the DTMC model, comprising a \emph{process layer}, an \emph{attacks layer}, and a \emph{mitigations layer}. This layered structure reflects the physical components, process, and informational interdependencies among the ballots affected by attacks, mitigations, and recourse actions. Furthermore, the DTMC allows for a nuanced understanding of the impact of policy implementations on the VBM process. In this section, we describe these three layers and discuss VBM performance. Although exact details of VBM operations may vary slightly according to a particular municipality in the US, the overall VBM process is similar between municipalities. Therefore, our high-level model of the VBM process effectively captures the fundamental dynamics and can be used to identify insights under different settings.

\subsection{ Process Layer}
\label{ssec: Process Layer}
First, we introduce seven process states that capture the movement of ballots in the process layer and define the operation of the VBM system. Each of these process states is also a state in the DTMC, and the states represent the change in physical location of the ballot over time. The voter requests a ballot in state $I$, the election office then fulfills this request in state $II$ and mails an unmarked ballot to the voter. \added{We implicitly assume that all ballots are sent by request to voters by mail, although in some US states, a voter may pick up an absentee ballot in person at the Board of Elections and return it later.} In state $III$ the unmarked ballot is handled by the United States Postal Service (USPS) and is in transit to the voter. In state $IV$ the unmarked ballot reaches the voter, however, it is up to the voter when the marked ballot is filled out and returned. In state $V$ the voter selects the method of return for the marked ballot. Standard operation for returning a ballot is a transition to state $VI$ via the USPS. 
Alternatively, the voter could return the ballot by drop box, if drop boxes are available, \added{which allows the associated ballot to bypass stage VI}. The final process state $VII$ occurs when the marked ballot reaches the election office and is held until processing on election day. The ballot process is summarized by the following states:

\begin{itemize}
    \item[-] \textbf{I:} Ballot requested by voter.
    \item[-] \textbf{II:} Unmarked ballot sent from election office.
    \item[-] \textbf{III:} Unmarked ballot in-transit via USPS.
    \item[-] \textbf{IV:} Voter marks ballot.
    \item[-] \textbf{V:} Voter returns marked ballot via USPS or drop box.
    \item[-] \textbf{VI:} Marked ballot in-transit via USPS.
    \item[-] \textbf{VII:} Marked ballot processed at election office and held to be counted on election day.
\end{itemize}

\subsection{ Final Ballot States }\label{ssec:finalstates}
At the end of the VBM process, ballots move to a final absorbing state that reflects the ballot status and is used to evaluate election performance. The final states capture all possible final states the voter ballot can take on at the end of the process, and they are defined to be mutually exclusive.
The voter ballots can be counted (C) or not counted (NC) by the election officials. There are several reasons a ballot may not be counted, and not all result from attacks to the process. Some ballots are not counted simply because they arrive at the election office after the required date for processing (L), which is typically an election day. \added{Ballots are not counted if they are missing the voter's signature, the signature does not match the signature on file, or the ballot is not properly sealed in the envelope or secrecy sleeve.}

In traditional VBM processes the only ballot states that are recorded are not counted and counted ballots, and all ballots are assumed to be unaltered (U). Some ballots can be altered (A) by masquerade attacks. Since an altered ballot may not be observable, altered ballots could be counted or not counted. Additionally, ballots may be lost, and all lost ballots are not counted and listed as not returned (NR).

Together, a ballot can be in one of the six following final ballot states at the end of the time horizon considered in the DTMC:

\begin{itemize}
    \item[-] \textbf{(C,U):} Voter ballot is received on-time unaltered in-transit, accepted, and counted on election day.

    \item[-] \textbf{(NC,U):} Voter ballot arrives on-time and unaltered, however it is rejected as incomplete.
    
    \item[-] \textbf{(NC,L):} Voter ballot is returned and \added{either fails to arrive by the election day or is postmarked after the election day (per relevant voting law).}
    
    \item[-] \textbf{(C,A):} Voter ballot is maliciously altered, accepted, and counted.
    
    \item[-] \textbf{(NC,A):} Voter ballot is maliciously altered, and it is rejected and not counted.

    \item[-] \textbf{(NC,NR):} Voter does not return the ballot, includes ballots lost in transit and ballots never delivered \cite{wec2020report}, \added{which typically triggers the provisional voting process.}
\end{itemize}

These final ballot states allow us to quantify the impact of various risks on an election. Note that the preferred ballot status is for ballots to reach the (C,U) state.

\subsection{Attacks Layer}
The attacks layer is comprised of active vulnerabilities to the VBM process and can reflect different attacker goals, including changing the outcome of an election or eroding trust in political systems. Vulnerabilities in the VBM system are documented in attack trees by the US Election Assistance Commission \cite{EAC_2009} and have been expanded upon since the 2020 General Election \cite{ScalaProcessMap}. The terminal nodes of these attack trees represent ``access'' points to additional vulnerabilities within the process. For this reason, we model each of these terminal leaf nodes of the attack trees \added{and they are labeled numerically with a prefix of ``X,'' using the notation from Scala et al.~\cite{ScalaProcessMap}. For example $X13$ represents the terminal attack node associated with the malicious loss of a ballot. } 
Attack trees, a valuable tool in cybersecurity, help assess threats to a system \cite{attacktrees}. They depict potential vulnerabilities and attacker paths in a Boolean logic tree structure, where the root node represents the ultimate goal of an attacker. Attack trees use a combination of Boolean AND($\times$) and OR($+$) logic gates to trace paths from terminal attack nodes back to the root node. 
Figure \ref{fig:X{14}X13} shows an example of an attack tree that visualizes two attacks corresponding to the malicious loss ($X{13}$) and accidental loss ($X{14}$) of a ballot \cite{EAC_2009}.
In Figure \ref{fig:X{14}X13}, terminal attack nodes $X{13}$ and $X{14}$ form part of an attack path leading to a successful insider attack, represented by root node 1 with only OR-gates along the paths. These attack paths offer a static view of threats within the VBM process, with the terminal leaf nodes $X{13}$ and $X{14}$ serving as entry points. 

\begin{figure}[htp]
  \centering
  \includegraphics[width=0.3\linewidth]{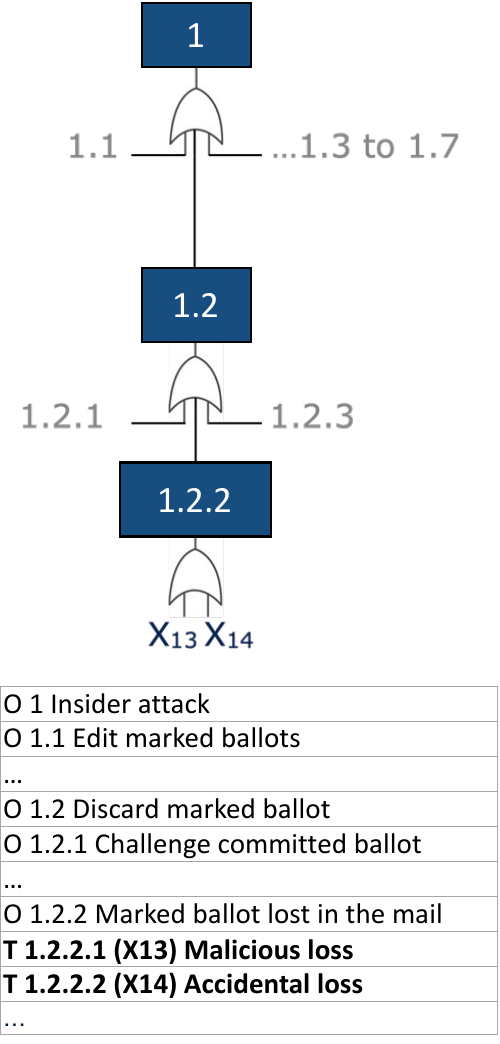}
  \caption{Example of a portion of the VBM attack tree \cite{EAC_2009}  }
  \label{fig:X{14}X13}
\end{figure}

\citet{iise_cah} explore the impact of attacks on the VBM process, focusing on how each attack could affect voter ballots in terms of being lost, late, or maliciously altered. Expanding upon this, here we classify attacks not only based on their impact on ballots but also on the VBM process as a whole. Terminal attack nodes are classified into three categories: \emph{fail rate increase}, where attacks change the \added{relative likelihood of} an undesirable outcome; \emph{ballot altering}, which involves direct tampering with the voter ballot; and \emph{process-altering}, where attacks cause voter ballots to deviate from the standard physical VBM process. This broader perspective allows for a more comprehensive understanding of the attacks' implications.

Any attack that affects the \added{relative likelihood} an outcome, e.g., a transition in the DTMC model, is considered to be a \emph{fail-rate} increase. These attacks can be non-malicious, e.g., caused by voter errors when filling out a ballot. As a result, the election office has a higher rejection rate for returned ballots, resulting in uncounted votes. However, these attacks can also be malicious, e.g., caused by a bad actor at the election office erroneously failing ballots.
For example, consider a malicious insider attack ($X9$) in which a VBM ballot is erroneously deemed insufficiently filled out (errant failed signature). If the malicious attack occurs, it increases the \added{relative likelihood} that a completed ballot is rejected. 

\added{Changing a marked ballot alters the information in the original ballot, which} we consider a ``ballot-altering attack.'' Masquerade attacks involve a bad actor maliciously altering a ballot. These attacks are more complex in that they create alternate ballot states in the DTMC that are needed to reflect the ballot's change in composition. 
For example, a masquerade attack in which a bad actor votes on behalf of someone in a central housing situation requires several steps. \added{In this type of attack, which is believed to be extremely rare,} the malicious actor must first register to vote on behalf of the original voter, intercept their mail, mark the ballot, and return the altered ballot to the VBM process. We designated this type of attack as ballot modifying. 
It is important to note that modified ballots still have the potential to be counted. For this reason, there is an alternate path for the modified ballots in the DTMC. This is later illustrated in the DTMC model attack layer in Section \ref{sec:mathmodel}, which reflects the VBM process states in the process layer to indicate that the ballots of those states were maliciously altered during transit. 

In the VBM system, there are seven distinct states that an original ballot must pass through to be counted and remain unaltered (see Section \ref{ssec: Process Layer}). When someone attempts to disrupt this process, it is known as a \emph{process-altering} attack. For example, \added{for illustration purposes} suppose an election volunteer steals the marked ballot while it is in transit to the election office. 
This \added{unlikely} malicious attack would alter the ballot process by holding the ballot in an ``attack state'' outside of the normal stages of the election process until the voter is notified and uses a recourse action such as requesting a replacement ballot, or it is deemed not returned and is not counted. The notable implication of process-altering attacks is that they can delay or prevent ballots from reaching the election office.

Table \ref{tab:structural_attacks} lists the terminal attack nodes to the VBM system from \citet{ScalaProcessMap} \added{that require voter or election office recourse actions to be countered.} The first column lists the updated terminal attack nodes in the VBM attack tree, while the second column identifies the specific stage in the VBM process affected by each attack. The third column indicates the final state of the ballot if the attack is successful. Table \ref{tab:structural_attacks} delineates the intent and classification of each terminal node attack in the last two columns, offering insights into the nature of these threats. 
\added{The terminal attack notes in boldface in Table \ref{tab:structural_attacks} are considered to represent the most significant risks to VBM according to a detailed risk analysis by Scala et al.~\cite{scala2022evaluating}, and therefore, we model these terminal attack nodes in our approach. The model can be extended in a straightforward manner to include other terminal attack nodes.}

\input{table_threats2}
There are malicious and non-malicious attack types. Malicious attacks are targeted and have a high local impact for a limited period of time (e.g., a day) of being active. Conversely, non-malicious attacks (e.g., voter error and accidental loss) are accidental and could occur any time over the DTMC model time horizon. 
For example, in the 2020 General Election, some ballots were not counted due to voters failing to sign or bundle the ballots correctly (attack $X{67}$). This non-malicious attack results in ballots being rejected by the election office, thus inhibiting the VBM process if not countered. We express this attack as a \added{low-likelihood occurrence} for each ballot over the entire evaluation period.

\subsection{Mitigations Layer}
\label{sssec:modeling:mitigations}
The mitigations layer in the DTMC model incorporates processes and actions designed to counteract both malicious and non-malicious threats. A mitigation is a recourse action available to counter any attacks to ballots in the VBM process. 
We build on previously published models of VBM mitigations to identify those that counteract attacks on the VBM process \cite{iise_cah}.

The mitigation layer connects with both the process and attack layers. We model structural mitigations \added{in} both the physical pathways (arcs) and the transitions of a ballot's progression. \added{In the context of the analysis, these transition values may represent relative likelihoods or stress-test intensities rather than empirically derived probabilities, allowing us to evaluate potential risk exposure in the absence of historical data.} Table \ref{tab:Mitigations List} details the five structural mitigation mitigations (\(M3, M4, M5, M6, M7\)) that require specific responsive measures from voters to negate the effects of attacks. These mitigations derive from  prior work that condense countermeasures published by CISA into practical logic intertwined with the VBM process \cite{iise_cah,cisa2020_VBM_risk}, and \added{the mitigation numbering starts at 3 to be consistent with the original labels in \cite{iise_cah}. Other mitigations exist \cite{Gregorio2024,iise_cah}, however, we focus on the core set of mitigations that are in use during the voting process}. Table \ref{tab:Mitigations List} presents the mitigations  categorized according to their operational nature along with the controlling entity, election office or voter.

\input{table_mitigations_simple}

This framework necessitates a thorough understanding of the connections between terminal node attacks in the VBM process and the available mitigation strategies designed to counteract them. To facilitate this, we employ a framework that examines the impact of terminal node attacks on the VBM process and incorporates recourse actions available to counter these attacks. Table \ref{tab:MitigationCoverage} illustrates the relationship between attacks and mitigations. In this table, an ``O'' indicates that a mitigation strategy can counter the threat without impacting the ballot processing time. In contrast, a ``D'' signifies that while mitigation can counter the threat, it introduces a delay in the VBM process. For example, the scenario of a lost ballot ($X{13}$), a type of malicious attack, causes a delay resulting from the mitigation logic that includes the time required to notify the voter about their missing ballot followed by the time it takes for the voter to decide on an appropriate recourse action, such as M4 or M6. This scenario exemplifies how Table \ref{tab:MitigationCoverage} summarizes key linkages in the DTMC model introduced in the next section. Consistent with the attacks in Table \ref{tab:structural_attacks}, terminal attack nodes listed in bold are incorporated into the DTMC model and are included in the computational results. 

\input{MitigationsCoverage}

\subsection{Modeling Assumptions}
\textcolor{black}{To ensure transparency in the application of the DTMC framework, we consolidate several key assumptions here. First, the model assumes that the high-level VBM process remains functionally consistent across different US municipalities. Second, the transition values used in the stress-test scenarios considered later in this paper represent relative likelihoods of risk exposure rather than strictly empirical probabilities, reflecting the inherent difficulty in estimating the frequency of rare security incidents. Third, the model treats the final ballot states (e.g., Counted, Unaltered (C,U)) as mutually exclusive and absorbing, allowing for the quantification of the final outcome distribution. Finally, while non-malicious risks are modeled as persistent baseline threats, malicious attacks are assumed to have a localized duration of typically one day to assess the impact to targeted attacks and identify specific times of heightened vulnerability within the election cycle.}

\section{DTMC Model} \label{sec:mathmodel}
In this section, we introduce a DTMC model of the VBM system based on a multi-layer configuration of the process, attacks, and mitigations layers. The DTMC model captures the stochastic movement of ballots through various stages from a ballot request to the counting of ballots. \added{It aggregates granular operational voting steps into functional steps to} analyze how different factors influence overall election performance.
A crucial aspect of the DTMC model is its ability to delineate the interaction between the terminal attack nodes, the mitigation layer, and the VBM process.

The time-inhomogeneous DTMC operates over a finite time horizon starting at time step \(t=1\) and continues until the final time step $T$.
The time between time steps is one day, with the state reflecting the system's state at the end of day $t$. 
Let \(t=1\) capture the earliest time election officials process requests for absentee ballots. The election is held at time step \(T-1\), under the assumption that ballots are not accepted after an election day. The model can easily be adapted to consider accepting ballots postmarked by election day by adding extra time steps.

\begin{figure*}[h]
  \centering
  \includegraphics[width=\linewidth]{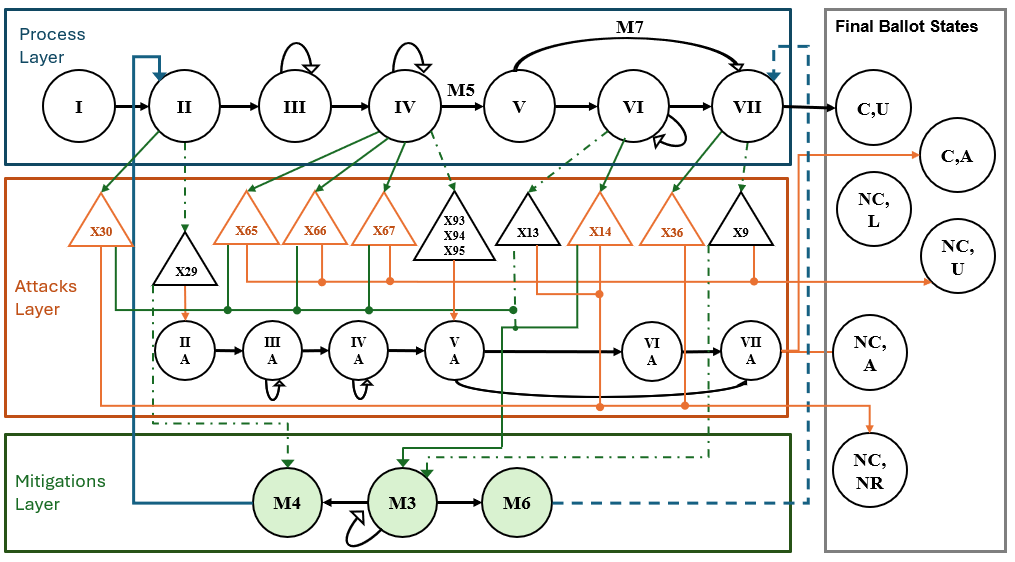}
  \caption{Layered Network for time intervals  \(1 \le t < T-1 \)}
  \label{fig:VBMLayered}
\end{figure*}

On election day, the arcs are re-positioned to move a ballot to its final post-election status in the last period, \(t=T\), to evaluate the performance measures.
Given a random process $V_t$ with $n$ finite states, the one-step transition probability of the process moving from state $i$ in time step $t$ to state $j$ in a single time step is
\begin{equation} \label{eq1}
    P_t(i,j) = P ( V_{t+1} =j | V_{t}=i ) \quad \forall i,j\in \{1,\cdots,n\}.
\end{equation}
We define a \added{state transition} matrix $P_t$ for each time step (\(t =1,2,\cdots, T-1\)) to calculate the ballot states at the end of each time step \cite{DTMCMath1}.
Figure \ref{fig:VBMLayered} illustrates the DTMC model state diagram for \(t<T-1\), showing the final ballot states along with three distinct layers: the process layer, the attack layer, and the mitigation layer. 
The DTMC model comprises 30 finite states \(S\in V_t\), categorized into six recurrent final ballot states and 24 transient states. 
Next, we describe the DTMC states and \added{transitions}, starting with the process layer and then adding the attack and mitigation layers.

The voting process initiates in the ``I'' state within the process layer, where voters begin the process by requesting a ballot. The ballot then navigates through various states in the process layer until it reaches one of the final ballot states. Table \ref{tab:structural_attacks} details the interactions and connections between these states in the VBM process layer and the attacks layer. 
In Figure \ref{fig:VBMLayered}, different shapes represent distinct Markov states of the VBM system. Triangles depict attacks while circles depict the physical states of the ballots. Triangles labeled in black denote malicious attacks, whereas those in gold represent non-malicious attacks. When an attack is active, ballots transition from the process layer state to one of the triangular attack states.  The ``mitigations layer''
encompasses recourse actions to counter attacks. Both \(M5\) and \(M7\) are illustrated in the process layer as these are \emph{process-altering} mitigations. Mitigation \(M5\)  affects the rate of ballot returns from voters, and \(M7\) enables voters to return their ballot in a drop box.

Additionally, Figure \ref{fig:VBMLayered} uses line styles to convey information about the arc \added{transitions} in the DTMC. Solid lines indicate transitions with non-zero probabilities at all steps, except at the final time step \(T-1\). In contrast, dotted and dashed lines represent transitions with non-zero \added{relative likelihoods} only at specific time steps, such as those associated with transient malicious attacks. The complexity of the model leads to overlapping lines in the figure; intersections marked by a dot signal a connection to the intersecting line.
In the model, masquerade attacks change a ballot's status and add a hidden attribute. Process layer states are mirrored in the attacks layer and labeled with an `A' in Figure \ref{fig:VBMLayered} to indicate the altered status of these ballots.
%
%
The DTMC model allows us to study VBM system performance as reflected by the final states of ballots over the course of an election cycle. We can determine the number of ballots that end up in the final desired state of ``Counted, Unaltered'' (C, U) and the other final ballot states.

\begin{figure*}[h]
  \centering
  \includegraphics[width=\linewidth]{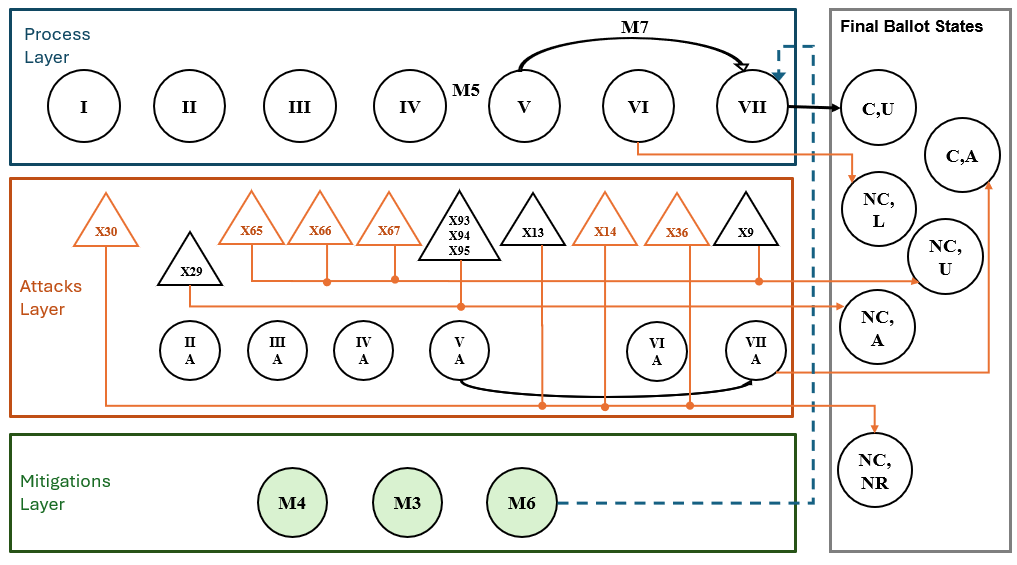}
  \caption{Layered Network for VBM on Election Day at time interval \( t\ge T-1\) }
  \label{fig:Pt0VBMLayered}
\end{figure*}

 At the end of the time horizon, the ballots move to the final ballot states. Figure \ref{fig:Pt0VBMLayered} shows the non-zero transition \added{relative likelihoods} on election day at time step \(T-1\). If ballots are not returned to the election office, they are ``not counted, not returned'' (NC, NR). These six final ballot states are recurrent in the DTMC, and all others are transient. If ballots are returned to the election office at \(t=T\)
they are ``not returned, late''(NC, L).

Next, we summarize the transitions. Let $P_t$ capture the \added{state transition} matrix immediately after the time steps $t=1,2,..., T-1$. Voters can request ballots at different times. Let $\beta_t$ capture the number of ballots requested at time $t=1,2,..., T-1$, which reflects the distribution of times when voters request absentee ballots. For the ballots requested at time $t$, let $\alpha_t$ be the vector of the probability mass function for the starting states of the ballot, where all the ballots are initiated in state $I$, that is, $\alpha_t(I)= 1 $ and $\alpha_t(s) \ge 0$ for all other states $s \ne I$.
For a ballot requested at time $t$, we can compute the vector of state probabilities at the end of the time horizon $\omega_{t}$ at time step $T$ after all ballots have been evaluated and transitioned to one of the ballot status states:
\begin{equation}
    \omega_t = \alpha_t \, \prod_{t'=t}^{T-1} P_{t'}, t = 1,2,..., T-1.
\end{equation}
Then, we can compute the overall distribution of final ballot statuses as $\sum_{t=1}^{T-1}\beta_t \omega_t $ that reflects the expected number of ballots in each final ballot state. \added{Note that several U.S. states send ballots to all registered voters. The DTMC model can be adapted to this situation by setting the initial ballot request parameter at time 1 ($\beta_1$) to the total number of registered voters and setting $\beta_t = 0$ for $t>1$.}

\section{Wisconsin 2020 General Election Case Study}\label{sec:casestudy}

We conduct a case study based on data from the 2020 General Election in Milwaukee County, Wisconsin. In 2020 Milwaukee County, the largest county in Wisconsin, had 939,489 residents and 478 voting wards \cite{census2019,mke2012,polling2020}. Milwaukee County election officials used multiple absentee voting mechanisms during the 2020 General Election, including VBM with designated drop boxes and in-person absentee voting. The state of Wisconsin allowed absentee voting with ``no excuses'' \added{(i.e., all voters are eligible for requesting an absentee ballot)} and sent voter reminders to return ballots \cite{NCSL}. At that time, the county had 550,132 registered voters and experienced a turnout of 83.67 percent \cite{wec2020registration}. There are no known malicious attacks on VBM in Milwaukee County in the 2020 General Election, although there were two convictions elsewhere in Wisconsin associated with independent instances of VBM election fraud, each affecting a single ballot \cite{heritagefraud}. \added{Certain low-volume attacks, such as the theft of a single blank ballot from a mailbox, are inherently difficult to detect and may be indistinguishable from benign process errors (e.g., accidental postal loss). As a result, we consider hypothetical worst-case threat scenarios when modeling the malicious  attacks to provide insight into the defensive measures could reduce risk or threat exposure. }

We assembled a data set for Milwaukee County using detailed, publicly reported information regarding absentee voting rates \cite{wec_absentee_nov2020,mec_nov2020_2020}. The Wisconsin Elections Commission reports a daily tally of the following events for each county prior to the election \cite{WIabsentee}:
\begin{enumerate}
\item absentee ballot requests,
\item ballots mailed out by an election office,
\item absentee ballots returned by mail, and
\item in-person absentee ballots cast.
\end{enumerate}

Using this data, we determined the daily number of absentee ballots requested, $\beta_t$, for $t=1, 2, \ldots, T-1$, with 94.1 percent of these ballots being returned. 
The procedures and recourse actions available to voters during the General Election on November 3, 2020 are represented in Table \ref{tab:Mitigations List} as mitigations. 
All requested ballots correspond to the starting state of the DTMC state $I$. Additional data were obtained from the MIT Election Data + Science Lab and the United States Office of the Inspector General \cite{mit_election_data__science_lab_voting_2021,USPS_2020}. The case study begins with establishing transition \added{likelihoods} for each arc in the process for every time interval $t=1,2,...,T$ in the DTMC model.

\subsection{Time}
In the VBM process, ballot states of the DTMC model transitions on a daily basis. The case study begins on September 17, 2020 ($t=1$), the first day ballots are mailed to voters, and ends after election day, November 4, 2020 ($T=49$). The DTMC transition \added{likelihoods} are time-dependent in accounting for procedure changes over the VBM timeline. For example, some mitigations are not available at all times due to VBM procedure, and malicious attacks occur at various times. However, many of the transition \added{likelihoods} are homogeneous across time intervals. As a result, we partition the time horizon into four segments with \added{transitions} that are time-homogeneous within an interval, \added{not including those associated with malicious attacks that are described in Section \ref{sec:results}.}

In the first interval, absentee voting opens and ballots are mailed to voters. The second interval begins when in-person absentee voting opens for the election. To reflect this procedure, mitigation $M6$ (related to in-person absentee voting) is unavailable during the first time interval. In the third time interval, ballots are no longer mailed to voters requesting absentee voting. During this interval, there is no longer a connection between the process states $I$ and $II$. Instead, the connection from state $I$ changes to $M6$, indicating that voters can only vote in-person absentee during this period. The fourth-time interval is election day and the days following, during which ballots are processed. Ballots that have not reached the USPS or the election office by election day are not counted. The exact final destination of the ballots depends on their location on election day. The model configuration in Figure \ref{fig:Pt0VBMLayered} outlines all arc connections from the voter requesting a ballot, $I$ of the process layer to one of the ``final ballot states.''   
\label{sec:time intervals listed}
In summary, the case study defines the time intervals as follows: 
\begin{itemize}
    \item interval 1 starts on September 17, 2020 ($t=1$ to $t=34$),
    \item interval 2 starts on October 21, 2020 ($t=35$ to $t=42$),
    \item interval 3 starts on October 29, 2020 ($t=43$ to $t=47$), and
    \item interval 4 starts on November 3, 2020 ($t=48$ to $t=49$). 
\end{itemize}

\begin{center}

\input{table_inputs}

\end{center}
%
\subsection{\added{Transitions} }
We define the \added{state transition} matrices as follows. We first define the \added{transitions} for the \emph{baseline model} that only considers non-malicious attacks. Non-malicious attacks are represented as consistent threats with time-inhomogenous \added{relative likelihoods} of occurrence. Later, we consider malicious attacks with a duration of one day. 

Table \ref{tab: ARC-Calibration} outlines the transition \added{relative likelihoods} for each arc active in the baseline model. The first and second columns define the two DTMC states associated with a non-zero transition \added{relative likelihood}. The next column provides a brief functional description of the ballot state transition that the arc represents. The columns labeled Interval 1, 2 and 3 represent the values of associated transition \added{relative likelihood} across time intervals 1, 2, and 3. We omit interval 4, since these transition \added{relative likelihoods} are $0$ or $1$ with arcs illustrated in Figure \ref{fig:Pt0VBMLayered}. The justification column reports a reference for each transition value and denotes which arcs were adjusted through calibration. The arc configuration aligns with the model shown in Figure \ref{fig:VBMLayered}.

We discuss several transitions.
The transition \added{relative likelihood} from process state $III$ to $IV$ is $0.938$, which reflects the proportion of election mail that is processed on-time as reported by the USPS Inspector General \cite{USPS_2020}. The rest of the ballots remain in state $III$. 
The transition from state $IV$ to $V$ reflects the return of completed ballots by voters (that is, the voter rate of ballot return). This transition captures the proportion of voters who return the marked ballot within a day of receiving it in the mail, which is determined as a part of the model calibration since this value is not directly recorded. 
\citet{mit_election_data__science_lab_voting_2021} provide the values for utilization of drop boxes over USPS to return ballots. Transitions from state $V$ to state $VII$, which occur with a \added{relative likelihood} of $0.515$, represents the proportion of ballots returned by drop boxes \cite{mit_election_data__science_lab_voting_2021}. 
Transitions at the end of the time interval move ballots to their final ballot states with \added{relative likelihood} 1.0. These probabilities can be gleaned from Figure \ref{fig:Pt0VBMLayered}. 
All other transition \added{relative likelihoods} are zero. 
\label{secc:results:modelcalibration}
We calibrated the DTMC using historical data from the WEC 2020 General Election report \cite{wec2020report}. This involved adjusting the voter rate of ballot return, the arc \added{transition likelihood} from node $IV$ to $V$, to align the output of (C,U) and (NC,U) ballots with the observed ballot return rate seen in aggregate by the state of Wisconsin. Following calibration, we validate the model's predictive accuracy by testing it against independent data validation points for Milwaukee County. This ensures blind prediction input data such that our model fits historical data and ensures the model is capable of making accurate predictions in varying scenarios.

\subsection{Calibration}
We calibrated the model using Wisconsin state-level data to align the recorded  empirical values of the ballots returned on various days with the calculated number of ballots counted and unaltered (\(C,U\)) final ballot state of the DTMC model. To accomplish this, we set the \emph{voter rate of ballot return} to reflect the influence of mitigation \(M5\) (reminders); the arc transition  from node \(IV\) to node \(V\) reflects the voter rate of ballots to return in intervals 2 and 3. 

\begin{figure}[h]
  \centering \includegraphics[width=0.7\linewidth]{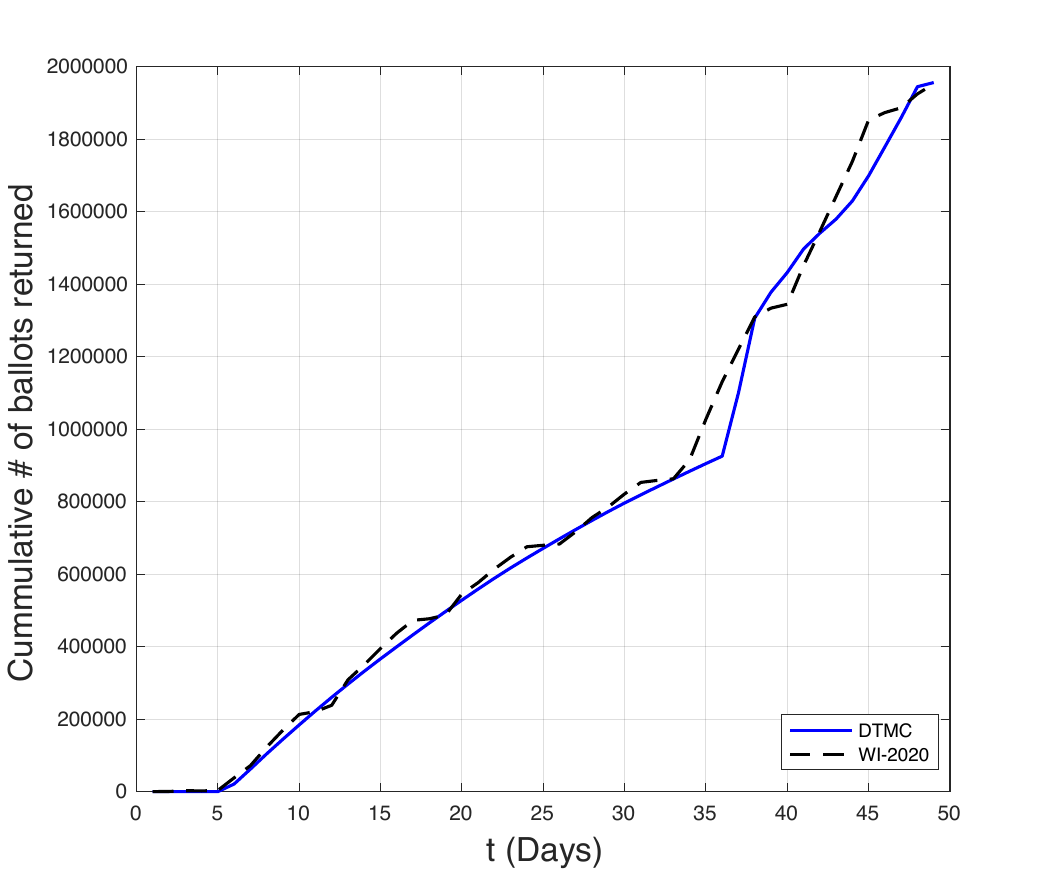}
  \caption{Comparison of recorded and modeled returned ballots for the state of Wisconsin for the 2020 General Election. The graph shows daily ballots returned in blue and the DTMC model baseline for daily ballots returned as a dashed line. }
  \label{fig:WIcalibration}
\end{figure}

Figure \ref{fig:WIcalibration} illustrates the cumulative number of ballots returned by day (solid line) compared to the expected number of ballots returned in the DTMC model (dotted line) using the calibrated return rates. The comparison of actual and modeled returned ballots for Milwaukee County is aligned with actual daily ballots returned (blue) and the baseline of the DTMC model for daily ballots returned (dashed). Note that all mitigations are not active in all time intervals. For example, the mitigation \(M5\) for reminders to return ballots was not implemented until time $t=35$ (interval 2). Figure \ref{fig:WIcalibration} shows that the modeled baseline model aligns closely with real-world data.

\input{table_baseline}
We then consider the mitigations of the model to determine a starting point for policies active in the 2020 Wisconsin General Election. We study the availability of mitigations M3, M4, M5, M6, and M7. Mitigation strength refers to the effect of each mitigation on the magnitude of the associated arc transitions. \added{Therefore, the mitigation strength values are relative likelihood values.}  Table \ref{tab:BaseMitigation} lists the inputs associated with mitigation strength in the baseline model.
Wisconsin did not implement automatic ballot notification in 2020, when voters were able to manually view ballot notifications through \url{myvote.wi.gov}. \added{We set mitigation \(M3\) strength to 0.0265 to align the model's baseline rejection rate with the empirical value of} (NC,U) ballots.
Next, we set mitigation \(M4\) to 0.90 to represent the ability of voters to request replacement ballots. We set mitigation \(M5\) to 0.74 to represent the sent and advertised reminders to return ballots in the late second interval. In Table \ref{tab: ARC-Calibration} we see the arc from node \(IV\) to node \(V\) increases in interval 2 to account for the increased rate of ballot return caused by implementation of mitigation \(M5\). Next, we set mitigation \(M6\) to 0.40 to represent the rate voters choose to submit an absentee ballot in person. Finally, we set the strength of \(M7\) to the average use in the state of Wisconsin \cite{mit_election_data__science_lab_voting_2021}. This results in a mean absolute deviation of 0.25-percent from the observed number of (NC,U) ballots.

The calibration of the baseline DTMC model also requires consideration of non-malicious attacks; we model as part of normal system operations. The \emph{ballot rejection rate} of the model is associated with the arc transition \added{relative likelihood} from node $IV$ to \( (X{65} + X{67} + X{68})\). It correlates directly with the number of ballots rejected by the election office due to voter error. We set the ballot rejection rate to 0.000162 to closely match the observed number of returned ballots that were unaltered and not counted (NC,U). 
Next, we vary the impact of the remaining non-malicious attacks, representing the ballots that never returned to the election office. Recall terminal attack node \(X14\) represents the accidental loss of ballots in the mail. The terminal attack node \(X30\) is the misaddressing of ballots to the voter, and \(X36\) represents the accidental loss of ballots in the election office. These non-malicious attacks all result in the same penalty, and the ballots impacted are considered not counted and not returned (NC,NR). Unlike the ballot rejection rate of the election office, there is a lack of evidence to support the different magnitudes for the non-malicious attacks \(X{14}, X{30}, X{36}\). These non-malicious attack strength values are equivalent for all periods $t=1,..., T-1$  \cite{wec2020analysis}. Therefore, we set the terminal attacks \(X{14}, X{30}, X{36}\) to equal strength. As a result, the following arcs have the same transition \added{relative likelihood} of 0.0343: arc from $IV$ to $X14$; arc from $II$ to $X30$; arc from $VII$ to $X36$. These arcs result in a mean absolute deviation of 0.14-percent from the observed (C,U) ballots.

\subsection{Validation}
\input{table_validation}
To validate the DTMC, we align the model outputs with five validation points from the 2020 WEC report scaled to Milwaukee County by ratio of ballots returned \cite{wec2020report}.
The validation points are as follows: 
\begin{itemize}
    \item  \textbf{v1}: Captures all returned and counted ballots. WEC reports these values for Milwaukee County on each day of the election cycle.
    \item  \textbf{v2}: Captures all ballots sent and not returned to the election office. WEC reports these values for Milwaukee County on each day of the election cycle.
    \item  \textbf{v3}: Reflects the ballots returned to the election office and rejected due to non-malicious attacks ($X{65}$, $X{66}$, $X{67}$). WEC reports the state level for the rejection rate of ballots; population ratios are used to scale to Milwaukee County. 
    \item  \textbf{v4}: Reflects all ballots deemed late as of day 48 of the election cycle. Again, WEC reports the state level for late ballots; population ratios are used to scale for Milwaukee County. 
    \item  \textbf{v5}: Captures the sum of all sent ballots. WEC reports these values for Milwaukee County on each day of the election cycle.
\end{itemize}

Table \ref{tab:validation} summarizes the validation points and relates them to the final ballot states of the model. 
The last two columns of Table \ref{tab:validation} show the number of ballots under each validation point, comparing real world ballot counts ``WEC ballot counts
scaled to Milwaukee'' and the DTMC model ballot counts. The validation process, aligning the DTMC model with real-world data from the WEC report, reinforces the model's reliability in reflecting absentee voting behaviors and provides valuable insights into vote-by-mail dynamics in Wisconsin.The last column of Table \ref{tab:validation} shows the DTMC model output values for the given validation points.
Comparative analysis aligned the model's final ballot states with the WEC report's metrics. Once calibrated and validated the model, most transitions were established for all time intervals. Milwaukee County data points are then used to establish a baseline model that can be used to consider multiple scenarios.

\section{Computational Results}\label{sec:results}

This section discusses the baseline DTMC model results for the Milwaukee County case study. The model is coded and implemented in Matlab, and the code and data are publicly available \cite{CH-github}. 
We then study the impact of several types of malicious attacks on the established baseline. We report the expected number of ballots in each of the final ballot states---rounded to the nearest ballot---under multiple scenarios. 

\subsection{Baseline model}

We first analyze the VBM system under normal conditions with only non-malicious attacks in the baseline model introduced in Section \ref{sec:casestudy}. Non-malicious attacks (see Table \ref{tab: ARC-Calibration}) are included in the baseline model, since they are an uncontrolled, continuous part of the VBM process. Additionally, we perform a sensitivity analysis of four of the non-malicious attacks by varying their corresponding \added{relative likelihoods} of occurring.

Table \ref{tab:MKEbase} summarizes the expected number of ballots in each final ballot state: counted and unaltered (C,U), not counted and unaltered (NC,U), correctly counted but altered (C,A), not counted due to late arrival (NC,L), not counted due to not being returned (NC,NR), and not counted and altered (NC,A).
Table \ref{tab:MKEbase} also reports a sensitivity analysis that varies the election office rejection rate and three other non-malicious attack transitions used in the model. 
The election office rejection rate varies based on ballots returned and rejected by the election office on election day due to non-malicious attacks $X{65}, X{66},$ and $X{67}$, according to subject matter expert testimony \cite{WIBarryBTestimony}.
The non-malicious attack strength of attacks $X{14}, X{30},$ and $X{36}$ represent ballots not returned due to either not being delivered to the voter or being discarded at the election office before counting the valid returned ballots on election day \cite{wec2020report}.

\input{table_MKEbase}

The top row of Table \ref{tab:MKEbase} represents the baseline model, and it reports the expected number of ballots in each final ballot state. Each subsequent row represents a different scenario with varying attack parameters. The incremental increase of the rejection rate associated with the arc transition from node $IV$ to \( (X{65} + X{67} + X{68})\) by a \added{value} of 0.00001  increases the number of (NC,U) ballots by 6-percent (35 total ballots), suggesting that efforts to reduce the rejection rate can decrease the number of ballots that are not counted.
Higher attack strengths associated with \(X{14}, X{30}, X{36}\) increase the number of ballots not returned (NC,NR). For example, when the attach strength increases from $0.0343$ to $0.0576$, the number of (NC,NR) ballots increase by $10,074$. 
Additionally, we note that the election office rejection rate ($X65+X66+X67$) for Milwaukee County is substantially lower than that of the other non-malicious attacks. 
These insights underscore the need for additional attention to unreturned (NC,NR) ballots in the VBM process and maintaining minimal non-malicious attack strengths to ensure that ballots are returned and counted.

\subsection{Malicious attack scenarios given baseline} 
\label{ssec:MaliciousAttacks}
Next, we introduce malicious attacks to the baseline. We focus on malicious attacks $X9, X13,$ and $X29$ \added{to illustrate the three attack types---fail rate increase, ballot modifying, and process altering.} 
Malicious attacks are modeled to last a single day during the 49-day time horizon to reflect feasible real-world election interference events, and we vary the attack strength of each attack to consider different scenarios. \added{We consider worst-case attack strengths to stress-test the model.}
Malicious attacks therefore, modify the transition  matrix (\(P_{t^*}\)) for a single day associated with the attack $t^*$. This approach allows us to assess the impact of attack timings and the mitigation strategies for countering attacks.

\subsubsection{X9: Malicious attack to challenge the signature of a valid ballot}
Malicious \(X9\) attack captures bad actors in an election office erroneously rejecting marked ballots by challenging their signatures. This scenario, with a base arc \added{transition} of 0.055, is set to affect roughly one of 19 precincts in Milwaukee. Referencing Figure \ref{fig:VBMLayered}, malicious attack \(X9\) impacts process state \(VII\), where ballots are collected and verified by the election office. On a specific day, this could lead to a significant number of ballots being improperly discarded.
\input{table_X9vtm36}
Table \ref{tab:X9vtm36} details the impact of the \(X9\) attack, showing variations in the expected number of ballots across the final ballot states under varying attack strengths and days. The top section reports the scenario of a medium-strength attack on different days. The following four sections of Table \ref{tab:X9vtm36} selects a day in the interval and varies the attack strength. These comparisons show the impact of attack strength versus attack day. 
We pay particular attention to ballot state (NC,U), since attack $X9$ causes ballots to not reach this state. Recall that there are 559 ballots in the (NC,U) final ballot state (see Table \ref{tab:MKEbase}).
Table \ref{tab:X9vtm36} reports how the attack day and attack strength affects the ballot outcomes. In the top section we find Day 38 to be particularly critical, in that 2,415 ballots end in the (NC,U) final ballot state. This increase in affected ballots is caused by increased ballot inflow due to active mitigations \(M5\) and \(M6\), initiated on day 36 (October 21, 2020). This finding underscores a vulnerability when attack strength is minimal but timed with peak ballot presence, highlighting critical periods when attacks have a larger scope of impact.

\subsubsection{X13: Malicious attack discard ballot in transit via USPS}
\input{table_X13vtm38}
Malicious $X{13}$ attack represents bad actors in a postal office who discard completed ballots at the sorting or storage point. 
Referring back to Figure \ref{fig:VBMLayered}, we see that malicious attack $X{13}$ directly impacts process state $VI$ where the voter returns the completed ballot via USPS. In Milwaukee County, a large-scale attack affects one of the 36 USPS offices for a day \cite{MilwData}. Consequently, there is a \added{relative likelihood} of 0.033 that ballots in process state \(VI\) are impacted by the $X{13}$ attack. 

Table \ref{tab:X13vtm38} summarizes the impact of attack \(X13\), showing variations in the expected number of ballots across different final states under varying attack strengths and days. The top section reports the scenario of a medium strength attack on different days. The following two sections of Table \ref{tab:X13vtm38} selects a day in the interval and varies the attack strength to determine the impact of attack strength versus attack day. 
Attack \(X{13}\) affects the number of ballots that are not returned, leading to an increase in the (NC,NR) final ballot state. The highest number of (NC,NR) ballots occurs when there is a malicious attack on day 37 of the election cycle. There are 21,219 ballots expected to not be returned (NC, NR), an 848 additional ballots not returned when compared to the baseline.

\subsubsection{X29: Masquerade Attacks, Ballot Modifying}
\label{ssec:ballotalter}
We then consider the malicious attack $X{29}$, which alters voters' original ballots and directly impacts process states $II$. 
The consequence of this attack leads to altered ballots that are both counted and uncounted (final ballot states (C,A) and (NC,A)). 
Since there exists no known reference of an executed masquerade attack, we assume that there is an equal \added{likelihood} that the altered ballot are counted or uncounted.
The manipulation from malicious attack $X{29}$ targets the early stages of ballot circulation, redirecting ballots from the process layer state $II$ to the attack layer state $II, A$. 

\input{table_X29vtm5}
Table \ref{tab:X29vtm5} summarizes the results associated with various days and strengths of attack \(X{29}\). The first column lists the scenarios considered, and the next two columns report the attack day and strength associated with different scenarios. The remaining columns report the expected number of ballots in each of the final ballot states. The final ballot states (C, U) and (C, A) reflect the impact of attack $X{29}$. Table \ref{tab:X29vtm5} shows the variations in the intensity of the attack during two distinct intervals, namely intervals 1 and 2. Intervals 3 and 4 are not studied, since ballots are not mailed to voters in these intervals. The attack day greatly affects the impact of the attack on the VBM process, affecting up to 664 ballots on a day 7 attack. Attacks carried out in the initial stages of the election have a large impact, since many ballots are mailed at these times. 

Overall, these results reveals a clear trend: initiating an attack in a high-demand phase of the election cycle results in a higher number of affected ballots. 

\subsubsection{Attack timing}
\label{ssec:TimingAttacks}
We examine the impact of attack timing on three malicious attack scenarios: \(X9\), \(X13\), and \(X29\). To do so, we vary the day of each attack from Day 7 to Day 49 and evaluate the change in the expected number of returned ballots that are counted and unaltered (C,U) compared to the baseline (see Table~\ref{ssec:MaliciousAttacks}).

\begin{figure}[ht]
  \centering
  \includegraphics[width=0.8\linewidth]{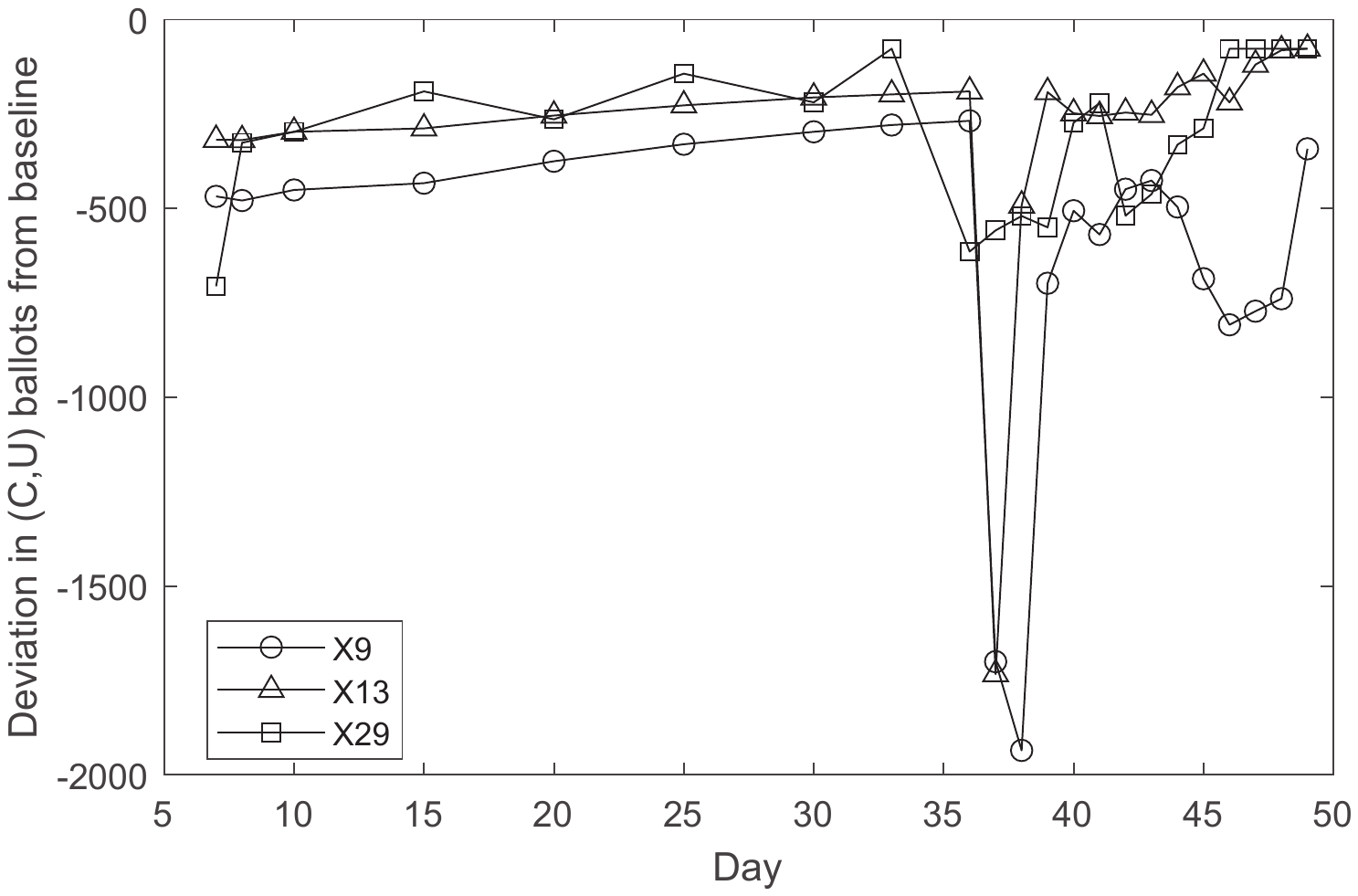}
  \caption{Deviation in counted and unaltered (C, U) ballots under three moderate malicious attack scenarios (\(X9\), \(X13\), and \(X29\)) launched on different days. 
  }
  \label{fig:timing}
\end{figure}

Figure \ref{fig:timing} illustrates the deviations in the expected number of (C,U) ballots associated with each attack scenario as a function of the day of the attack. All values are negative, indicating that each attack reduces the number of ballots in the preferred (C,U) state. 
Attack \(X9\), which models the false rejection of valid ballots at the election office (process state VII), experiences its largest deviation in (C,U) ballots when the attack is initiated on day 38. This attack peak aligns with a surge of ballots routed to the election office \added{by mail} via early voting (M6), which begins on day 36 and introduces a direct arc from process state II to state VII, in addition to those being returned by mail or drop box.
Similarly, attack \(X13\), representing the loss of ballots in USPS transit, peaks on day 37 in terms of its deviation in the expected number of (C,U) ballots. This is due to ballot return reminders (M5) beginning on Day 36. This \added{prompts voters to return their ballots and} results in an increase in the flow of ballots in transit, which results in more ballots being available to be maliciously stolen by malicious actors. \added{However, attack \(X13\) affects few ballots in the week prior to election day, when most ballots have already been returned.}

Attack \(X29\), a masquerade attack involving stolen or pre-marked ballots (state II), experiences its largest deviation in (C,U) ballots earlier in the election cycle on day 7, which results in 706 fewer (C,U) ballots. This occurs due to the initial surge in ballot issuance and early returns at this time. A secondary deviation of 614 (C,U) ballots occurs on day 36, coinciding with an increase in voters returning ballots due to automatic ballot reminders (M5). These results indicate that attack timing--—specifically its alignment with the flow of ballots--—is crucial for determining the consequences of an attack. The consequences of an attack corresponds to its interaction with process dynamics during periods of high ballot flow.  \added{As with \(X13\), \(X29\) affects few ballots in the week prior to election day.}

\subsection{Worst-Case Scenario Modeling}
\label{ssec:policy_mitigations_test}
We build on the attack scenarios in Section \ref{ssec:TimingAttacks} to create a ``worst-case scenario'' to evaluate various policy implementations in the VBM process under severe conditions as well as the impact on mitigations. \textcolor{black}{The findings presented in this section are contingent upon the assumed strengths of these mitigations and the strengths of the associated attacks. }

The worst-case scenario involves the simultaneous execution of three malicious attacks within the 49-day election cycle. Figure \ref{fig:timing} illustrates the highest impact for each attack as $X{9}$ on day 38, $X{13}$ on day 37, and $X{29}$ on day 7. The following arc \added{transitions} are all set to 0.10 on these days: arc from node $IV$ to $X{13}$ on day 37; arc from node $II$ to $X{29}$ on day 7; arc from node $VII$ to $X{9}$.
We also consider a sensitivity analysis that evaluates the impact of mitigation strength and associated strength against the worst-case attack scenario. 
Referencing Table \ref{tab:MitigationCoverage} in Section \ref{sssec:modeling:mitigations}, four mitigations counter the attacks in the worst-case scenario (see Table \ref{tab:MitigationCoverage}):
\begin{itemize}
    \item $X{9}$ is countered by automatic ballot notifications M3, replacement ballots M4, and early voting M6,
    \item $X{13}$ is countered by automatic ballot notifications M3, replacement ballots M4, early voting M6, and drop boxes M7,
    \item $X{29}$ is countered by  automatic ballot notifications M3 and replacement ballots M4.
\end{itemize}

Table \ref{tab:worstcasescenario} reports final ballot states associated with various mitigation scenarios, reflected by the first five columns. Then, it outlines the expected number of ballots in each final ballot state in the remaining six columns.

We first vary drop box ($M7$) availability  from its base value of $0.520$ using an availability range from $0.1$ to $0.95$. We do not consider drop boxes to have zero availability, since voters can return their absentee ballots directly to their local municipality election administration office. The remaining mitigations ($M6, M4, M3$) vary in availability, ranging from $0.01$ to $0.95$ to capture a wide range of operating conditions based on current federal laws and reports of mitigation effectiveness \cite{bipartisan_policy_center_2021}.

The first row of Table \ref{tab:worstcasescenario} shows the impact of the worst-case malicious attack scenario on the DTMC model for Milwaukee County. We see 8,950 fewer ballots counted and unaltered (C,U) than the baseline. Further, we see 1,208 altered ballots in final ballot states (C,A) and (NC,A). The number of ballots that are not counted increases significantly, with 3,593 (NC,U) ballots compared to 559 in the baseline.
The results in Table \ref{tab:worstcasescenario} show a substantial impact in the performance of the VBM process when compared to the baseline results with no malicious attacks in Table \ref{tab:MKEbase}. Next, we vary mitigation strength to find which mitigations might be effective in countering the affects of the worst-case scenario. The goal is to align the worst-case with the baseline through only modifying mitigation strengths.  

Next, we vary mitigation strength. Drop box $M7$ availability significantly impacts the number of ballots in the final ballot states. Increasing the availability of $M7$ to $0.95$ increases the counted ballots (C,U) to 324,867, which is higher the 324,792 (C,U) ballots in the baseline scenario without malicious attacks. Reducing $M7$ availability to less than 0.10 results in 17,776 (NC,U) ballots, the highest across all worst-case scenarios. However, M7 does not counter masquerade attacks to reduce the number of altered ballots (C,A) or (NC,A).

The following eight rows in Table \ref{tab:worstcasescenario} examine the availability of early voting $M6$. Changes in early voting $M6$ availability has substantially less impact on final ballot states than drop boxes $M7$ availability. Similar to drop boxes $M7$, the availability of early voting $M6$ fails to counter masquerade attacks by reducing the number of altered ballots (C,A) or (NC,A). 

Next, we examine varying mitigations automatic ballot notifications $M3$ and replacement ballots $M4$ simultaneously to reflect their interdependency. We find that deploying automatic ballot notifications $M3$ with the same availability as replacement ballots$M4$, at $0.90$, results in 325,584 counted (C,U) ballots, an increase of 9,742 ballots as compared to the baseline, worst-case performance in the first row of Table \ref{tab:worstcasescenario}. An automatic ballot notifications $M3$ availability of 0.90  results in 62 expected (C,A) ballots. 
Lastly, since automatic ballot notifications $M3$ and early voting $M6$ can function interdependently, we examine the impact of simultaneously varying automatic ballot notifications $M3$ and early voting $M6$. When deployed at 0.90 availability, we observe 325,485 (C,U) ballots, similar to the previous observations. Additionally, we observe fewer ballots not returned, with 17,243 (NC,NR) ballots compared with 20,371 (NC,NR) ballots in the Milwaukee baseline without malicious attacks. We conclude that \added{early} absentee voting gives voters a powerful recourse option to counter malicious and non-malicious attacks. 

\input{table_worstcasescenario}
\begin{figure}[htp]
\centering
  \includegraphics[width=0.6\linewidth]{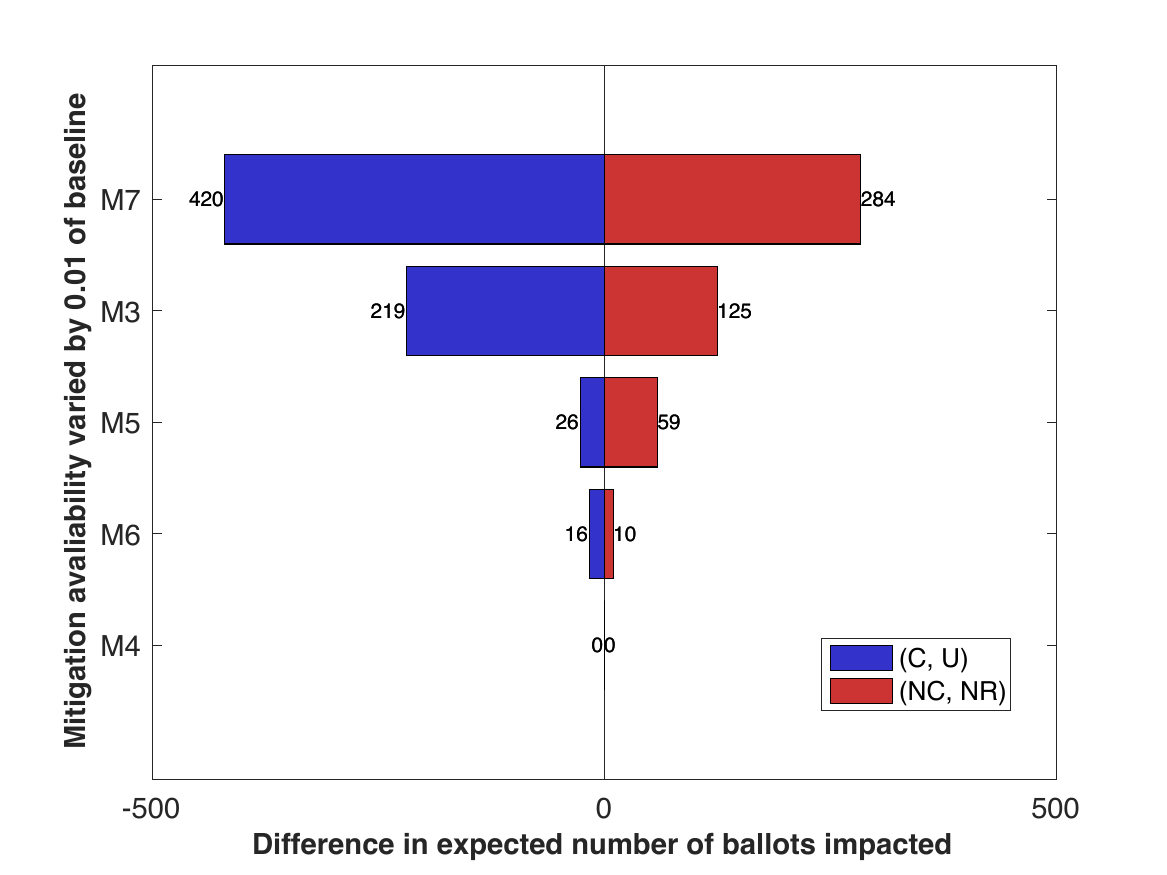}
  \caption{Results of a sensitivity analysis that evaluates the impact of mitigations M3, M4, M5, M6, and M7 on the DTMC model's final ballot outputs in the presence of multiple malicious attacks. We compare the changes in the expected number of (C, U) and (NC, NR) ballots to the baseline.}
  \label{fig: tornado}
\end{figure}

Next, we study the sensitivity of the DTMC model mitigation strength. The two final ballot states we study are the desired final ballot state (C,U) and the unreturned ballots (NC,NR) since these two final ballot states are the most impacted by mitigations. However, other ballot states are impacted by mitigations to a lesser degree.  We perform a one-way sensitivity analysis by changing each mitigation strength by $\pm 0.01$ from its baseline value (see Table \ref{tab:Mitigations List}). 
Figure \ref{fig: tornado} shows a tornado graph illustration of the results. 
The blue half of Figure \ref{fig: tornado} on the left represents the difference in the number of correctly counted and unaltered ballots when the mitigation strength decreases. The red half is the difference in the number of ballots that are not counted and not returned, which occurs when mitigation strength increases. 
The tornado graph shows that the DTMC model is most sensitive to drop boxes $M7$ and automatic ballot notifications $M3$. In contrast, replacement ballots $M4$ and early voting $M6$ (independently) do not significantly affect the outputs in a significant manner. However, their impact can be larger when their availability is considered within the context of other mitigations (see Table~\ref{tab:worstcasescenario}). These results demonstrate that modeling attack timing in conjunction with mitigation availability provides crucial insight into the consequence of each attack.
\textcolor{black}{Our findings depend on the assumed strengths of these mitigations and the strengths of the associated attacks. The number of affected ballots in real settings would also depend on voter behavior and the operational reliability of election processes.}

\section{Conclusions}\label{sec:conclusion}
VBM process and policy are rapidly evolving in the United States. Our research presents a quantitative tool designed to \added{stress-test Vote-by-Mail systems.}  We introduce a discrete-time Markov Chain modeling framework, which we apply to a case study based on the 2020 General Election in Milwaukee County to evaluate the security and effectiveness of VBM processes. \added{The case study analysis frames threats as hypothetical scenarios to compare the relative impact of defensive mitigations. Considering ``worst-case'' conditions with threat actors allows us to identify which mitigations have the greatest effect, independent of the actual (and historically low) base rate of attack.}
Our findings highlight the critical role of dynamic mathematical modeling in computational risk analysis for VBM systems. This layered approach effectively organizes the complex interactions among cyber, physical, and human elements in VBM, providing insight into VBM vulnerabilities. Our analysis, which leverages various ballot outcomes and real-world data, 
\textcolor{black}{is framed as a stress-test of systemic vulnerabilities rather than a literal forecast of election outcomes. By evaluating various hypothetical, worst-case scenarios, the model serves to identify aspects of the VBM infrastructure that are most sensitive to disruption.}

\added{Our analysis suggests that} singular mitigations, such as drop boxes, can bolster the VBM system, while other mitigations must be combined to reduce risk. A crucial observation is that automatic ballot notifications (dependent on replacement ballots or early voting) is the only mitigation capable of effectively countering ballot-altering malicious attacks. The variation of each mitigation policy set changes the VBM process's resilience to the worst-case scenario. These changes mark quantifiable metrics for evaluating VBM security and highlight the need for an integrated and adaptive mitigation approach to ensure the integrity of the VBM process. 

Our research highlights the fact that risk in VBM process is dynamic and must be managed through a systems approach. We note that different malicious attacks may have different scopes of impact based on the timing of the attacks. Our findings suggest that mitigations that provide voters with information and recourse actions are crucial for VBM security. In particular, drop boxes enable voters to cast a second ballot when their first ballot is affected by a malicious or non-malicious attack. 

Additionally, automatic ballot tracking can significantly affect VBM performance. Both drop boxes and automatic ballot tracking provide broad protection against attacks. Two notable examples of ballot tracking programs that are currently available are BallotTrax and BallotScout. BallotTrax allows voters to track the status of their mail-in ballots, providing real-time updates when ballots are collected, received, and accepted by election offices \cite{BalloTrax}. Similarly, BallotScout integrates with USPS to provide voters with detailed information on their ballots' whereabouts throughout the election process \cite{BallotScout}.

There are several limitations of our study that provide insight into topics for future research. First, this analysis considers malicious attacks that occur on a single-day and is limited by a lack of historical data regarding malicious attacks. Future research could diversify attack scenarios and incorporate a broader range of data to enhance the model's sensitivity to resource constraints and varying electoral processes. Second, future research could account for resource constraints, system congestion, and \added{additional mitigations such as enhanced voter education and ballot design}. Third, we did not examine the likelihood of attack scenarios altering election outcomes. Successful attacks on voting systems compromise the integrity of elections and undermine public trust in political systems, regardless of whether they change the outcomes. Although not all attacks aim to change election results, this remains a significant concern that future research could investigate.

\added{Finally, there is significant epistemic uncertainty in this application area. It assumes that the defined states and transitions capture the full spectrum of possible ballot movements and threats. However, this structure cannot inherently predict ``unknown unknowns,'' such as entirely new attack vectors or rapid legislative shifts that fundamentally alter the voting process. The results presented in this paper should be interpreted as a risk assessment of the current security landscape. The model is designed to be extensible; therefore, the underlying Markov chain structure must be updated as new threats or process change emerge.}

\section*{Acknowledgments}
This work was in part funded by the National Science Foundation Award 2000986. The views and conclusions contained in this document are those of the authors and should not be interpreted as necessarily representing the official policies, either expressed or implied, of the National Science Foundation.
\added{The authors would like to thank the anonymous reviewers and the associate editor, whose suggestions for improvement led to a substantially improved manuscript.}

\bibliographystyle{abbrvnat}
\begin{singlespace}
\bibliography{VBMref} 
\end{singlespace}

\end{document}

%% file: table_threats2.tex
\begin{table*}[hbt!]
\caption{Modeled VBM terminal attack nodes and their attributes}
\label{tab:structural_attacks}
\resizebox{\textwidth}{!}{
\begin{tabular}{|@{}l|cccc|}
\toprule
 & Terminal attack node & Final ballot state & & \\
Attack tree & linkage to Process Layer & for successful attack & Intent & Classification \\ \hline
T 1.2.1.1.1 (X8) Judge misinterprets rule                                  & VII                                                                                                                      & NC, U                                                                                                               & Malicious                            & Fail rate increase                           \\ 
\textbf{T 1.2.1.1.2 (X9) Errant failed signature}                                   & VII                                                                                                                      & NC, U                                                                                                               & Malicious                            & Fail rate increase                           \\ 
T 1.2.1.2.1 (X10) Challenge signature                                      & VII                                                                                                                      & NC, U                                                                                                               & Malicious                            & Fail rate increase                           \\ 
T 1.2.1.2.2 (X11) Challenge postmark                                       & VII                                                                                                                      & NC, U                                                                                                               & Malicious                            & Fail rate increase                           \\ 
T 1.2.1.2.3 (X12) Challenge intent                                         & VII                                                                                                                      & NC, U                                                                                                               & Malicious                            & Fail rate increase                           \\ 
\textbf{T 1.2.2.1 (X13) Malicious loss }                                           & VI                                                                                                                       & NC, NR                                                                                                           & Malicious                            & Process altering                             \\ 
\textbf{T 1.2.2.2 (X14) Accidental loss}                                           & VI                                                                                                                       & NC, NR                                                                                                           & Non-malicious                        & Process altering                             \\ 
T 1.5.1.1 (X28) Fail to stuff envelope                                     & II                                                                                                                       & NC,A                                                                                                                & Malicious                            & Ballot modifying                             \\ 
\textbf{T 1.5.1.2 (X29) Send wrong or pre marked ballot}                         & II                                                                                                                       & C,A                                                                                                                 & Malicious                            & Ballot modifying                             \\ 
\textbf{T 1.5.1.3 (X30) Mis-address envelope (to   voter)}                         & II                                                                                                                       & NC,NR& Non-malicious& Process altering                             \\ 
\textbf{T 1.5.3.1 (X36) accidentally lost in the mail room}               & VII                                                                                                                      & NC, NR                                                                                                           & Non-malicious                        & Process altering                             \\ 
T 1.5.3.2 (X37) Mailbox attack                                             & VI                                                                                                                       & NC, NR                                                                                                           & Malicious                            & Process altering                             \\ 
T 1.7.2 (X84) Vote denied                                                  & VII                                                                                                                      & NC,U                                                                                                                & Malicious                            & Process altering                             \\ 
T 2.3.1 (X43) Identify target residents                                    & IV                                                                                                                       & C,A                                                                                                                 & Malicious                            & Ballot modifying                             \\ 
T 2.3.2 (X44) Register them                                                & IV                                                                                                                       & C,A                                                                                                                 & Malicious                            & Ballot modifying                             \\ 
T 2.3.3 (X45) Intercept, mark, and return   their ballot                   & IV                                                                                                                       & C,A                                                                                                                 & Malicious                            & Ballot modifying                             \\ 
T 2.3.4.1 (X46) Register as the voter                                      & IV                                                                                                                       & C,A                                                                                                                 & Malicious                            & Ballot modifying                             \\ 
T 2.3.4.2 (X47) Forge the signature                                        & IV                                                                                                                       & C,A                                                                                                                 & Malicious                            & Ballot modifying                             \\ 
T 2.4.1 (X48) Identify target                                              & IV                                                                                                                       & C,A                                                                                                                 & Malicious                            & Ballot modifying                             \\ 
T 2.4.2 (X49) Steal blank ballot from   mailbox                            & IV                                                                                                                       & C,A                                                                                                                 & Malicious                            & Ballot modifying                             \\ 
T 2.4.3 (X50) Receive, mark, return their   ballots                        & IV                                                                                                                       & C,A                                                                                                                 & Malicious                            & Ballot modifying                             \\ 
T 2.4.4.1 (X51) Register as the voter                                      & IV                                                                                                                       & C,A                                                                                                                 & Malicious                            & Ballot modifying                             \\ 
T 2.4.4.2 (X52) Forge the signature                                        & IV                                                                                                                       & C,A                                                                                                                 & Malicious                            & Ballot modifying                             \\ 
    T 2.5 (X53) Malicious “messenger ballots”   & II                                                                                                                       & C,A                                                                                                                 & Malicious                            & Ballot modifying                             \\ 
\textbf{T 2.8.1 (X93) Steal blank ballot from mailbox}                            & IV                                                                                                                       & C,A                                                                                                                 & Malicious                            & Ballot modifying                             \\ 
\textbf{T 2.8.2 (X94) Mark and return their ballot}                              & IV                                                                                                                       & C,A                                                                                                                 & Malicious                            & Ballot modifying                             \\ 
\textbf{T 2.8.3 (X95) Defeat signature check }                                      & IV                                                                                                                       & C,A                                                                                                                 & Malicious                            & Ballot modifying                             \\ 
\textbf{T 4.1.1 (X65) Failure to sign correctly }                                  & IV                                                                                                                       & NC,U                                                                                                                & Non-malicious                        & Process altering                             \\ 
\textbf{T 4.1.2 (X66) Signature mismatch }                                        & IV                                                                                                                       & NC,U                                                                                                                & Non-malicious                        & Process altering                             \\ 
\textbf{T 4.1.3 (X67) Failure to bundle correctly}                                  & IV                                                                                                                       & NC,U                                                                                                                & Non-malicious                        & Process altering                             \\ 
\textbf{T 4.1.4 (X68) Failure to meet time   requirements}                          & IV                                                                                                                       & NC, L                                                                                                            & Non-malicious                        & Process altering                             \\ \bottomrule
\end{tabular}%
}
\end{table*}

%% file: table_mitigations_simple.tex
\begin{table}[htbp]
\centering
\caption{Mitigations available for the VBM process \cite{iise_cah} }
\resizebox{\linewidth}{!}{%
\begin{tabular}{|>{\raggedright\arraybackslash}m{0.25\linewidth}|m{0.1\linewidth}|m{0.4\linewidth}|m{0.2\linewidth}|}
\hline
\textbf{Mitigation name} & \textbf{Label} & \textbf{Mitigation description} & \textbf{Controlling entity}  \\ 
\hline
Automatic ballot notifications &\textbf{M3} & The ability of a voter to attain the status of their ballot. This has a high probability if automatic notifications are provided via BalloTrax/Ballot Scout.& Election office  \\ 
\hline

Replacement ballots &\textbf{M4} & Replacement ballot package request & Voter \\ 
\hline
Automatic ballot reminders &\textbf{M5} & Notify voter to send ballot back before deadline & Election office  \\ 
\hline
Early voting &\textbf{M6} & In-person absentee voting & Voter  \\ 
\hline
Drop boxes &\textbf{M7} & Return ballot via drop box & Voter  \\ 
\hline
\end{tabular}%
}
\label{tab:Mitigations List}
\end{table}

%% file: MitigationsCoverage.tex
\begin{table*}[hbp!]
\caption{Linkages between terminal attack nodes and mitigations }
\label{tab:MitigationCoverage}
\resizebox{\textwidth}{!}{%
\begin{tabular}{|lccccc|}
\hline
\multicolumn{1}{|c|}{\textbf{\begin{tabular}[c]{@{}c@{}}Terminal   Attack Nodes of \\      VBM Attack Tree\end{tabular}}} &
  \multicolumn{1}{c|}{\textbf{\begin{tabular}[c]{@{}c@{}}M3\\      Automated \\ notification of \\ ballot status\end{tabular}}} &
  \multicolumn{1}{c|}{\textbf{\begin{tabular}[c]{@{}c@{}}M4\\      Replacement\\      ballot voter\\  request\end{tabular}}} &
  \multicolumn{1}{c|}{\textbf{\begin{tabular}[c]{@{}c@{}}M5\\      Notify voters\\  to send ballot \\ back earlier\end{tabular}}} &
  \multicolumn{1}{c|}{\textbf{\begin{tabular}[c]{@{}c@{}}M6\\      In-person \\ absentee voting\end{tabular}}} &
  \multicolumn{1}{c|}{\textbf{\begin{tabular}[c]{@{}c@{}}M7\\      Return ballot \\ via drop boxes\end{tabular}}}  \\ \hline
 
T 1.2.1.1.1 (X8) Judge misinterprets rule &
  D &
  D &
   &
  D &
   \\
\textbf{T 1.2.1.1.2 (X9)   Errant failed signature} &
  \textbf{D} &
  \textbf{D} &
  \textbf{} &
  \textbf{D} &
  \textbf{} \\
 
T 1.2.1.2.1 (X10) Challenge signature &
  D &
  D &
   &
  D &
   \\
T 1.2.1.2.2 (X11)   Challenge postmark &
  D &
  D &
   &
  D &
   \\
 
T 1.2.1.2.3 (X12) Challenge intent &
  D &
  D &
   &
  D &
   \\
\textbf{T 1.2.2.1 (X13)   Malicious loss} &
  \textbf{D} &
  \textbf{D} &
  \textbf{} &
  \textbf{O} &
  \textbf{O} \\
 
\textbf{T 1.2.2.2 (X14) Accidental loss} &
  \textbf{D} &
  \textbf{D} &
  \textbf{} &
  \textbf{O} &
  \textbf{O} \\
T 1.5.1.1 (X28) Fail   to stuff envelope &
   &
  D &
   &
   &
   \\
 
\textbf{T 1.5.1.2 (X29) Send wrong or pre marked ballot} &
  \textbf{D} &
  \textbf{D} &
  \textbf{} &
  \textbf{} &
  \textbf{} \\
\textbf{T 1.5.1.3 (X30)   Mis-address envelope (to voter)} &
  \textbf{D} &
  \textbf{D} &
  \textbf{} &
  \textbf{O} &
  \textbf{} \\
 
\textbf{T 1.5.3.1 (X36) accidentally lost in the   mailroom} &
  \textbf{} &
  \textbf{} &
  \textbf{} &
  \textbf{} &
  \textbf{} \\
T 1.5.3.2 (X37)   Mailbox $\backslash$ Dropbox attack &
  D &
  D &
   &
  O &
   \\
 
T 1.7.2 (X84) Vote denied &
  D &
  D &
   &
   &
   \\
T 2.3.1 (X43) Identify   target residents &
  D &
  D &
   &
   &
   \\
 
T 2.3.2 (X44) Register them &
  D &
  D &
   &
   &
   \\
T 2.3.3 (X45)   Intercept, mark, and return their ballot &
  D &
  D &
   &
   &
   \\
 
T 2.3.4.1 (X46) Register as the voter &
  D &
  D &
   &
   &
   \\
T 2.3.4.2 (X47) Forge   the signature &
  D &
  D &
   &
   &
   \\
 
T 2.4.1 (X48) Identify target &
  D &
  D &
   &
  O &
   \\
T 2.4.2 (X49) Steal   blank ballot from mailbox &
  D &
  D &
   &
  O &
   \\
 
T 2.4.3 (X50) Receive, mark, return their ballots &
  D &
  D &
   &
  O &
   \\
T 2.4.4.1 (X51)   Register as the voter &
  D &
  D &
   &
   &
   \\
 
T 2.4.4.2 (X52) Forge the signature &
  D &
  D &
   &
   &
   \\
T 2.5 (X53) Malicious   “messenger ballots”  &
  D &
  D &
   &
   &
   \\
 
\textbf{T 2.8.1 (X93) Steal blank ballot from mailbox} &
  \textbf{D} &
  \textbf{D} &
   &
 &
  \\
\textbf{T 2.8.2 (X94) Mark and   return their ballot}&
  \textbf{D} &
  \textbf{D} &
   &
 &
 \\
 
\textbf{T 2.8.3 (X95) Defeat signature check} &
  \textbf{D}&
 \textbf{D} &
    &
   &
   \\
\textbf{T 4.1.1 (X65) Failure   to sign correctly} &
  \textbf{D} &
  \textbf{D} &
  \textbf{} &
  \textbf{} &
  \textbf{} \\
 
\textbf{T 4.1.2 (X66) Signature mismatch} &
  \textbf{D} &
  \textbf{D} &
  \textbf{} &
  \textbf{} &
  \textbf{} \\
\textbf{T 4.1.3 (X67) Failure   to bundle correctly} &
  \textbf{D} &
  \textbf{D} &
  \textbf{} &
  \textbf{} &
  \textbf{} \\
 
\textbf{T 4.1.4 (X68) Failure to meet   time requirements} &
  \textbf{} &
  \textbf{} &
  \textbf{O} &
  \textbf{O} &
  \textbf{} \\ \hline
\end{tabular}%
}
\end{table*}

%% file: table_inputs.tex


\begin{table*}[]
\centering
\caption{DTMC arc transition relative likelihoods}
\label{tab: ARC-Calibration}
 \begin{adjustbox}{width=1.15\textwidth,center}
\begin{tabular}{@{}|c|c|l|c|c|c|l|@{}}
\hline \textbf{From node} & \textbf{To node} & \textbf{Function} & \textbf{Interval 1} & \textbf{Interval 2} & \textbf{Interval 3} & \textbf{Justification} \\ \hline 
I &
  II &
  Voter requests ballot to election office &
  1 &
  1 &
  1 &   All voter requests show being \\ 
  &&&&&&received by election office \\ \hline 
II &
  III &
  Unmarked ballot sent from election office &
  1 &
  0.2 &
  0 &
  All ballot requests filled by the \\
  &&&&&&election office \\ \hline 
II &
  M6 &
  Ballots requests are deferred to in-person absentee &
  0 &
  0.8 &
  1 &
  WI policy for 2020 General Election \cite{WIabsentee} \\ \hline 
  II &
  X30 &
  Ballot envelope mis-addressed to voter (undelivered) &
  0.0343 &
  0.0343 &
  0.0343 &
  Calibration, \\
  &&&&&&\citeauthor{wec2020report} \cite{wec2020report} \\ \hline
III &
  IV &
  Unmarked ballot in transit via USPS &
  0.938 &
  0.938 &
  0.938 &
  Office of \citeauthor{USPS_2020} \cite{USPS_2020} \\ \hline
III &
  III &
  Unmarked ballot remains at USPS past one cycle &
  0.062 &
  0.062 &
  0.062 &
   Office of \citeauthor{USPS_2020} \cite{USPS_2020} \\ \hline
IV &
  IV &
  Voter waits one day to return ballot&
  0.947 &
  0.797 &
  0.947 &
  Calibration, \citeauthor{wec2020report} \cite{wec2020report} \\ \hline
IV &
  V &
  Voter returns marked ballot &
  0.05 &
  0.79&
  0.79&
  Calibration, \citeauthor{wec2020report} \cite{wec2020report} \\ \hline
IV &
  X65+X66+X67 &
  Ballot subject to non-malicious attack of voter error &
  0.000162&
  0.000162&
  0.000162&
  Calibration, \citeauthor{wec2020report} \cite{wec2020report}\\
  &&voter error&&&& \\ \hline
V &
  VI &
  Marked ballot returned via USPS &
  0.485 &
  0.485 &
  0.485 &
  \citeauthor{mit_election_data__science_lab_voting_2021} \cite{mit_election_data__science_lab_voting_2021} \\ \hline
V &
  VII &
  Marked ballot returned via Dropbox &
  0.515 &
  0.515 &
  0.515 &
  \citeauthor{mit_election_data__science_lab_voting_2021} \cite{mit_election_data__science_lab_voting_2021} \\ \hline
VI &
  VII &
  Marked ballot in transit and received  &
  0.938 &
  0.938 &
  0.938 &
 \\
 && at election office &&&&\citeauthor{USPS_2020} \cite{USPS_2020} \\ \hline
VI &
  X14 &
  Marked ballot lost at USPS, non-malicious  &
  0.0343 &
  0.0343 &
  0.0343 &
  Calibration, \\
  
  &&&&&&\citeauthor{wec2020report} \cite{wec2020report} \\ \hline
VI &
  VI &
  Marked ballot remains at USPS past one cycle &
  0.061 &
  0.061 &
  0.061 &
 Office of \\
 &&&&&&\citeauthor{USPS_2020} \cite{USPS_2020} \\ \hline
VII &
  (C,U) &
  Marked ballot processed at election office &
  0.966 &
  0.966 &
  0.966 &
  Ballots are be counted and\\ 
  &&&&&& unaltered if not subject to attack \\ \hline
VII &
  X36 &
  Marked ballot lost at election office, non-malicious &
  0.0343 &
  0.0343 &
  0.0343 &
  Calibration,\\
  &&&&&&\citeauthor{wec2020report} \cite{wec2020report} \\ \hline
M4 &
  II &
  Ballot replacement request &
  1 &
  1 &
  1 &
  Once mitigation is reached, \\ 
  &&&&&&request are be made to election office \\ \hline
M3 &
  M4 &
  Voter is notified ballot has problem and  &
  0.5 &
  0.33 &
  0.33 &
  Equal probability for active mitigations \\
  &&chooses   replacement&&&& \\ \hline
M3 &
  M6 &
  Voter is notified ballot has problem &
  0 &
  0.33 &
  0.33 &
  Equal probability for active mitigations \\
  &&  and chooses in-person vote &&&& \\ \hline
M3 &
  M3 &
  Voter is notified ballot has problem and &
  0.5 &
  0.34 &
  0.34 &
  Equal probability for active mitigations \\
  &&  chooses no recourse action &&&& \\ \hline
M6 &
  VII &
  In-person absentee voting &
  1 &
  1 &
  1 &
  Ballot successfully submitted \\ \hline
II-A &
  III-A &
  Modified ballot sent to the voter &
  1 &
  1 &
  1 &
  WI timeline for absentee voting  \cite{WIabsenteeDeadlines} \\ \hline
III-A &
  IV-A &
  Modified ballot received by the voter &
  0.9 &
  0.9 &
  0.9 &
  Altered ballot received by the voter \\ \hline
III-A &
  III-A &
  Modified ballot remains in transit beyond one cycle &
  0.1 &
  0.1 &
  0.1 &
  Attacked ballots movement low \\ \hline
IV-A &
  IV-A &
  Modified ballot is not filled out in one cycle &
  0.1 &
  0.1 &
  0.1 &
  Approximation (fast return for malicious intent) \\ \hline
IV-A &
  V-A &
  Modified ballot returned &
  0.9 &
  0.9 &
  0.9 &
  Approximation (fast return for malicious intent) \\ \hline
V-A &
  VI-A &
  Modified ballot returned via USPS &
  0.485 &
  0.485 &
  0.485 &
 \citeauthor{mit_election_data__science_lab_voting_2021} \cite{mit_election_data__science_lab_voting_2021} \\ \hline
V-A &
  VII-A &
  Modified ballot returned via Dropbox &
  0.515 &
  0.515 &
  0.515 &
  \citeauthor{mit_election_data__science_lab_voting_2021} \cite{mit_election_data__science_lab_voting_2021} \\ \hline
VI-A &
  VII-A &
  Modified ballot returned to election office on-time &
  1 &
  1 &
  1 &
 Office of \citeauthor{USPS_2020} \cite{USPS_2020} \\ \hline
VII-A &
  (C,A) &
  Modified ballot counted &
  0.5 &
  0.5 &
  0.5 &
  Equal probability  of being counted unless \\ 
  &&&&&&other mitigations are in   place \\ \hline
VII-A &
  (NC,A) &
  Modified ballot rejected at election office &
  0.5 &
  0.5 &
  0.5 &
  Equal probability  of being counted unless \\ 
  &&&&&& other mitigations are in   place \\ \hline
X14 &
  M3 &
  Voter notified of non-malicious attack of lost ballot &
  0.0265 &
  0.0265 &
  0.0265 &
  Calibration, \citeauthor{wec2020report} \cite{wec2020report}\\ \hline
X14 &
  (NC,NR)&
  Marked ballot lost in process &
  0.974 &
  0.974 &
  0.974 &
  Inverse of M3 availability \\ \hline
X36 &
  (NC, NR)&
  Ballot lost in election office &
  1 &
  1 &
  1 &
  No monitoring available, no recourse \\ \hline
X65+X66+X67 &
  M3 &
  Voter made aware of errors in ballot package &
  0.0265 &
  0.0265 &
  0.0265 &
  Calibration, \citeauthor{wec2020report} \cite{wec2020report}\\ \hline
X65/X67 &
  (NC,A) &
  Marked ballot rejected at election office &
  0.974 &
  0.974 &
  0.974 &
  Inverse of M3 availability \\ \hline
X93-X95 &
  V-A&
  Masquerade attack leading altered ballot path &
  0.10 &
  0.10 &
  0.10 &
  Test various strengths of attack on various days \\ \hline
X93-X95 &
  M4 &
  Mitigation M4 availability  &
  0.90 &
  0.90 &
  0.90 &
  Test various strengths of attack on various days \\ \hline
X9 &
  (NC,A) &
  Malicious challenge of ballot signature successful &
  0.974 &
  0.974 &
  0.974 &
  Test various strengths of attack on various days \\ \hline
X9 &
  M3 &
  Voter notified of failed ballot submission &
  0.0265 &
  0.0265 &
  0.0265 &
  Calibration, \citeauthor{wec2020report} \cite{wec2020report}\\ \hline
X29 &
  III-A &
  Malicious incomplete ballot sent to voter  &
  0.10 &
  0.10 &
  0.10 &
  Test various strengths of attack on various days \\
  && leading to altered   ballot path &&&& \\ \hline
X29 &
  M4 &
  Voter requests replacement ballot &
  0.90 &
  0.90 &
  0.90 &
  Test various strengths of attack on various days \\ \hline
\end{tabular}%
  \end{adjustbox}
\end{table*}

%% file: table_baseline.tex
\begin{table*}[]
\caption{Baseline values for mitigation deployment level}
\label{tab:BaseMitigation}\centering
\begin{tabular}{|l|c|l|}
\hline
\textbf{Mitigation name and label} &
  \textbf{Mitigation strength} &
  \textbf{Justification} \\ \hline
  & & Automatic voter notifications of ballot \\
Automatic ballot notifications (M3) &
  0.0265   & status were not implemented \\ \hline
& & Voters had option of requesting \\
Replacement ballots (M4) &
  0.900 &
   replacement ballots \\ \hline
Automatic ballot reminders (M5) &
  0.740&
  Region used ballot return reminders \\ \hline
Early voting (M6) &
  0.400&
  In-person absentee was implemented \\ \hline
&& Survey of the Performance of American \\
Drop boxes (M7) & 0.520 &  Elections Dataverse \cite{MIT_VBMa} \\ \hline\end{tabular} 

\end{table*}

%% file: table_validation.tex
\begin{table*}[]
\caption{Model validation using WI Election Commission (WEC) data }
\label{tab:validation}
\resizebox{\linewidth}{!}{
\begin{tabular}{|c|c|c|c|c|c|}
\hline
\textbf{\begin{tabular}[c]{@{}c@{}}Validation \\  point\end{tabular}} &
  \textbf{WEC ballot   status} &
  \textbf{Corresponding DTMC final ballot states} &
  \textbf{\begin{tabular}[c]{@{}c@{}}WEC ballot \\ counts\end{tabular}} &
  \multicolumn{1}{c|}{\textbf{\begin{tabular}[c]{@{}c@{}}WEC ballot counts  \\  scaled to \\ Milwaukee County \end{tabular}}} &
  \textbf{\begin{tabular}[c]{@{}c@{}}DTMC ballot \\ counts\end{tabular}} \\ \hline
v1 & Ballots   returned and counted        & (C,U)+(C,A)                                   & 1,969,274
& 324,896
& 324,792\\
v2 & Ballots   not returned               & (NC,   NR) & 85,586
& 20,349
& 20,371\\
v3 & Ballot   rejected by election office & (NC,U)+(NC,A)                                 & 3,225
& 532
& 559
\\
v4 & Ballot not returned before polls closed   & (NC,Late)                                     & 1,045
& 176
& 173
\\
& & County specific sum  &  &  & \\ 
v5 & All   sent ballots                   &  of all ballot states & 2,059,130 & 325,547& 325,351\\ \hline
\end{tabular}
}
\end{table*}

%% file: table_MKEbase.tex
\begin{table*}[]
\caption{Baseline DTMC model values and final ballot states  }
\label{tab:MKEbase}
 \begin{adjustbox}{width=\textwidth,center}
\begin{tabular}{|l|cc|cccccc|}
\hline
& Election office
& \( X{14}, X{30}, X{36}\)&
  \multicolumn{6}{c|}{\textbf{Final Ballot States (Expected number of ballots)}} \\ 
Scenario &
  rejection rate & attack strength & (C,U) & (NC,U) & (C,A) & (NC,L) &
(NC,NR) & (NC,A) \\ \hline
Baseline &
  0.000162&
  0.0343 &
  324,792&
  559&
  0&
  173&
  20,371&
  \multicolumn{1}{c|}{0} \\ \hline
Baseline that varies election office rejection rate &
  0.0000707&
  0.0343 &
  325,091&
  245&
  0&
  174&
  20,386&
  \multicolumn{1}{c|}{0} \\
 from node \(IV\) to \( (X{65} + X{67} + X{68})\) &
  0.000111&
  0.0343 &
   324,958 &
   385 &
  0&
   174 &
   20,379 &
  \multicolumn{1}{c|}{0} \\
\multicolumn{1}{|c|}{} &
  0.000172&
  0.0343 &
   324,759 &
   594 &
  0&
   173 &
   20,369 &
  \multicolumn{1}{c|}{0} \\
\multicolumn{1}{|c|}{} &
  0.000424&
  0.0343 &
   323,933 &
   1,464 &
  0&
   173 &
   20,325 &
  \multicolumn{1}{c|}{0} \\
\multicolumn{1}{|c|}{} &
  0.000818&
  0.0343 &
   322,653 &
   2,812 &
  0&
   172 &
   20,258 &
  \multicolumn{1}{c|}{0} \\ \hline
Baseline that varies non-malicious attack strength &
  0.000162&
  0.00202 &
   339,228 &
   559 &
  0&
   198 &
   5,911 &
  \multicolumn{1}{c|}{0} \\
 associated with $X14, X30, X36$ &
  0.000162&
  0.0192 &
   331,493 &
   559 &
  0&
   184 &
   13,660 &
  \multicolumn{1}{c|}{0} \\
\multicolumn{1}{|r|}{} &
  0.000162&
  0.0384 &
   323,025 &
   559 &
  0&
   171 &
   22,141 &
  \multicolumn{1}{c|}{0} \\
\multicolumn{1}{|r|}{} &
  0.000162&
  0.0576 &
   314,732 &
   560 &
  0&
   158 &
   30,445 &
  \multicolumn{1}{c|}{0} \\
\multicolumn{1}{|r|}{} &
  0.000162&
  0.0939 &
   299,469 &
   560 &
  0&
   140 &
   45,728 &
  \multicolumn{1}{c|}{0} \\ \hline
\end{tabular}
\end{adjustbox}
\end{table*}

%% file: table_X9vtm36.tex
\begin{table*}[]
\caption{Attack X9: The malicious attack of erroneously rejecting of ballots}
\label{tab:X9vtm36}
\resizebox{\textwidth}{!}{%
\begin{tabular}{|c|cc|cccccc|}
\hline
& Day of malicious
& Malicious &
  \multicolumn{6}{c|}{\textbf{Final Ballot States (Expected number of ballots)}} \\ 
Scenario &
   attack & attack strength & (C,U) & (NC,U) & (C,A) & (NC,L) &
(NC,NR) & (NC,A) \\ \hline
X9 active at medium strength varying attack days   &  10 &   0.055 
&  324,341 &  933 
& 0 &  173 &  20,449 
& 0 \\
                                                     &  30 &   0.055 
&  324,495 &  779 
& 0 &  173 &  20,449 
& 0 \\
                                                     &  36 &   0.055 
&  324,524 &  751 
& 0 &  173 &  20,449 
& 0 \\
                                                     &  37 &   0.055 
&  323,092 &  2,180 
& 0 &  173 &  20,451 
& 0 \\
                                                     &  38 &   0.055 
&  322,857 &  2,415 
& 0 &  173 &  20,451 
& 0 \\
                                                     &  39 &   0.055 
&  324,094 &  1,180 
& 0 &  173 &  20,449 
& 0 \\
                                                     &  40 &   0.055 
&  324,286 &  988 
& 0 &  173 &  20,449 
& 0 \\
                                                     &  45 &   0.055 
&  324,106 &  1,161 
& 0 &  173 &  20,457 
& 0 \\
                                                     &  48 &   0.055 
&  324,053 &  1,221 
& 0 &  173 &  20,449 
& 0 \\ \hline
Vary attack strength X9 with attacks in interval 1 &  10 &   0.050 
&  324,375 &  899 
& 0 &  173 &  20,449 
& 0 \\
                                                     &  10 &   0.055 
&  324,341 &  933 
& 0 &  173 &  20,449 
& 0 \\
                                                     &  10 &   0.075 
&  324,205 &  1,069 
& 0 &  173 &  20,450 
& 0 \\
                                                     &  10 &   0.100 
&  324,035 &  1,239 
& 0 &  173 &  20,450 
& 0 \\
                                                     &  30 &   0.050 
&  324,515 &  759 
& 0 &  173 &  20,449 
& 0 \\
                                                     &  30 &   0.055 
&  324,495 &  779 
& 0 &  173 &  20,449 
& 0 \\
                                                     &  30 &   0.075 
&  324,415 &  859 
& 0 &  173 &  20,449 
& 0 \\
                                                     &  30 &   0.100 
&  324,315 &  959 
& 0 &  173 &  20,449 
& 0 \\ \hline
Vary attack strength X9 with attacks in interval 2 &  40 &   0.050 
&  324,325 &  949 
& 0 &  173 &  20,449 
& 0 \\
                                                     &  40 &   0.055 
&  324,286 &  988 
& 0 &  173 &  20,449 
& 0 \\
                                                     &  40 &   0.075 
&  324,130 &  1,144 
& 0 &  173 &  20,449 
& 0 \\
                                                     &  40 &   0.100 
&  323,935 &  1,339 
& 0 &  173 &  20,450 
& 0 \\ \hline
Vary attack strength X9 with attacks in interval 3 &  45 &   0.050 
&  324,161 &  1,106 
& 0 &  173 &  20,456 
& 0 \\
                                                     &  45 &   0.055 
&  324,106 &  1,161 
& 0 &  173 &  20,457 
& 0 \\
                                                     &  45 &   0.075 
&  323,884 &  1,380 
& 0 &  173 &  20,460 
& 0 \\
                                                     &  45 &   0.100 
&  323,607 &  1,653 
& 0 &  173 &  20,464 
& 0 \\ \hline
Vary attack strength X9 with attacks in interval 4 &  48 &   0.050 
&  324,114 &  1,161 
& 0 &  173 &  20,449 
& 0 \\
                                                     &  48 &   0.055 
&  324,053 &  1,221 
& 0 &  173 &  20,449 
& 0 \\
                                                     &  48 &   0.075 
&  323,813 &  1,462 
& 0 &  173 &  20,449 
& 0 \\
                                                     &  48 &   0.100 &  323,512 &  1,763 & 0 &  173 &  20,449 & 0 \\ \hline
\end{tabular}%
}
\end{table*}

%% file: table_X13vtm38.tex
\begin{table*}[]
\caption{Attack X13: Ballot maliciously discarded in transit via USPS}
\label{tab:X13vtm38}
\resizebox{\textwidth}{!}{%
\begin{tabular}{|c|cc|cccccc|}
\hline
& Day of malicious
& Malicious &
  \multicolumn{6}{c|}{\textbf{Final Ballot States (Expected number of ballots)}} \\ 
Scenario &
   attack & attack strength & (C,U) & (NC,U) & (C,A) & (NC,L) &
(NC,NR) & (NC,A) \\ \hline
X13, are active at medium strength varying attack days& 10&  0.033&     324,495&     559
&     0&     173&     20,551&     0
\\
                                                      & 36&     0.033 &        324,602 &        559 
&     0&      173 &      20,502 &     0
\\
                                                      & 37&     0.033 &        323,059 &        559 
&     0&      173 &      21,219 &     0
\\
                                                      & 38&     0.033 &        324,300 &        559 
&     0&      173 &      20,642 &     0
\\
                                                      & 40&     0.033 &        324,544 &        559 
&     0&      173 &      20,529 &     0
\\
                                                      & 45&    0.033&       324,649&       559&     0&     173&     20,480&     0\\ \hline
Vary attack strength X13 with attacks in interval 1& 10&     0.025 &        324,521 &        559 &       0&        173 &        20,525 &     0
\\
                                                      & 10&     0.030 &        324,505 &        559 &       0&        173 &        20,541 &     0
\\
                                                      & 10&     0.040 &        324,473 &        559 &       0&        173 &        20,574 &     0
\\
                                                      & 10&     0.065 &        324,392 &        559 &       0&        173 &        20,654 &     0
\\
                                                      & 30&      0.025 &         324,601 &         559 &       0&         173 &         20,494 &     0
\\
                                                      & 30&      0.030 &         324,592 &         559 &       0&         173 &         20,503 &     0
\\
                                                      & 30&      0.040 &         324,573 &         559 &       0&         173 &         20,522 &     0
\\
                                                      & 30&   0.065 &      324,525 &      559 &     0&      173 &      20,570 &     0\\ \hline
Vary attack strength X13 with attacks in interval 2& 40&   0.025 &      324,564 &      559 &     0&      173 &      20,508 &     0
\\
                                                      & 40&   0.030 &      324,551 &      559 &     0&      173 &      20,521 &     0
\\
                                                      & 40&   0.040 &      324,526 &      559 &     0&      173 &      20,546 &     0
\\
                                                      & 40&   0.065 &      324,464 &      559 &     0&      173 &      20,609 &     0\\ \hline
\end{tabular}%
}
\end{table*}

%% file: table_X29vtm5.tex
\begin{table*}[]
\caption{Attack X29: Intercept and pre-mark ballot to voter}
\label{tab:X29vtm5}
\resizebox{\textwidth}{!}{%
\begin{tabular}{|c|cc|cccccc|}
\hline

& Day of malicious
& Malicious &
  \multicolumn{6}{c|}{\textbf{Final Ballot States (Expected number of ballots)}} \\ 
Scenario &
   attack & attack strength & (C,U) & (NC,U) & (C,A) & (NC,L) &
(NC,NR) & (NC,A) \\ \hline
X29 are active at medium strength varying attack days & 7  & 0.055&    324,086 &    558 &    332 &    173 &    20,416 &    332 
\\
                                                      & 8  & 0.055&    324,466 &    559 &    132 &    173 &    20,436 &    132 
\\
                                                      & 10 & 0.055&    324,495 &    559 &    116 &    173 &    20,437 &    116 
\\
                                                      & 20 & 0.055&    324,528 &    559 &    99 &    173 &    20,439 &    99 
\\
                                                      & 30 & 0.055&    324,573 &    559 &    75 &    173 &    20,441 &    75 
\\
                                                      & 40 & 0.055&    324,519 &    560 &    99 &    173 &    20,438 &    109 
\\ \hline
Vary attack strength X29 with attacks in interval 1   & 20 & 0.025&    324,630 &    559 &    45 &    173 &    20,444 &    45 
\\
                                                      & 20 & 0.050&    324,545 &    559 &    90 &    173 &    20,440 &    90 
\\
                                                      & 20 & 0.075&    324,460 &    559 &    135 &    173 &    20,435 &    135 
\\
                                                      & 20 & 0.100&    324,375 &    559 &    180 &    173 &    20,431 &    180 
\\
                                                      & 30 & 0.025&    324,651 &    559 &    34 &    173 &    20,445 &    34 
\\
                                                      & 30 & 0.050&    324,586 &    559 &    68 &    173 &    20,442 &    68 
\\
                                                      & 30 & 0.075&    324,521 &    559 &    102 &    173 &    20,439 &    102 
\\
                                                      & 30 & 0.100&    324,457 &    559 &    136 &    173 &    20,435 &    136 
\\ \hline
Vary attack strength X29 with attacks in interval 2   & 40 & 0.025&    324,626 &    559 &    45 &    173 &    20,444 &    50 
\\
                                                      & 40 & 0.050&    324,536 &    559 &    90 &    173 &    20,439 &    90 
\\
                                                      & 40 & 0.075&    324,447 &    559 &    134 &    172 &    20,435 &    149 
\\
                                                      & 40 & 0.100&    324,358 &    559 &    179 &    172 &    20,430 &    198 \\ \hline
\end{tabular}%
}
\end{table*}

%% file: table_worstcasescenario.tex
\begin{table*}[]
\caption{Policy performance for a worst-case election cycle} 
\label{tab:worstcasescenario}
\resizebox{\textwidth}{!}{%

\begin{tabular}{|l|cccc|cccccc|}
\hline

& M3 & M4 & M6 & M7&
  \multicolumn{6}{c|}{\textbf{Final Ballot States (Expected number of ballots)}} \\ 
Scenario &
   available & available & available & available & (C,U) & (NC,U) & (C,A) & (NC,L) &
(NC,NR) & (NC,A) \\ \hline
Milwaukee County baseline  performance for worst case & & & & & & & & & & \\  (X13, day 37) + (X29, day 7) +   (X9, day 38)  & 0.027& 0.900& 0.400& 0.520&  315,842 &  3,593 &  604 &  391 &  22,913 &  604 
\\ \hline
Vary drop box M7 availability for worst case                                             & 0.027& 0.900& 0.400& 0.100&  307,016 &  4,865 &  604 &  735 &  28,901 &  604 
\\
                                                                                         & 0.027& 0.900& 0.400& 0.200&  309,118 &  4,562 &  604 &  653 &  27,475 &  604 
\\
                                                                                         & 0.027& 0.900& 0.400& 0.520&  315,842 &  3,593 &  604 &  391 &  22,913 &  604 
\\
                                                                                         & 0.027& 0.900& 0.400& 0.750&  320,671 &  2,897 &  604 &  203 &  19,637 &  604 
\\
                                                                                         & 0.027& 0.900& 0.400& 0.850&  322,769 &  2,594 &  604 &  122 &  18,213 &  604 
\\
                                                                                         & 0.027& 0.900& 0.400& 0.950&  324,867 &  2,291 &  604 &  41 &  16,789 &  604 
\\ \hline
Vary in-person absentee M6 for worst case                                                & 0.027& 0.900& 0.010& 0.520&  316,160 &  3,198 &  604 &  423 &  23,099 &  604 
\\
                                                                                         & 0.027& 0.900& 0.100& 0.520&  316,086 &  3,289 &  604 &  416 &  23,056 &  604 
\\
                                                                                         & 0.027& 0.900& 0.200& 0.520&  316,005 &  3,390 &  604 &  407 &  23,009 &  604 
\\
                                                                                         & 0.027& 0.900& 0.300& 0.520&  315,923 &  3,492 &  604 &  399 &  22,961 &  604 
\\
                                                                                         & 0.027& 0.900& 0.400& 0.520&  315,842 &  3,593 &  604 &  391 &  22,913 &  604 
\\
                                                                                         & 0.027& 0.900& 0.500& 0.520&  315,760 &  3,694 &  604 &  382 &  22,865 &  604 
\\
                                                                                         & 0.027& 0.900& 0.750& 0.520&  315,557 &  3,948 &  604 &  362 &  22,745 &  604 
\\
                                                                                         & 0.027& 0.900& 0.950& 0.520&  315,393 &  4,150 &  604 &  345 &  22,650 &  604 
\\ \hline
Vary Ballot Trax/Scout implementation M3                     & 0.010& 0.010& 0.400& 0.520&  315,661 &  3,652 &  614 &  389 &  23,016 &  614 
\\
with mail-in availability M4                                                                                         & 0.027& 0.027& 0.400& 0.520&  315,842 &  3,593 &  604 &  391 &  22,913 &  604 
\\
                                                                                         & 0.050& 0.050& 0.400& 0.520&  316,099 &  3,508 &  590 &  393 &  22,766 &  590 
\\
                                                                                         & 0.400& 0.400& 0.400& 0.520&  319,963 &  2,236 &  372 &  423 &  20,558 &  372 
\\
                                                                                         & 0.500& 0.500& 0.400& 0.520&  321,078 &  1,868 &  310 &  432 &  19,921 &  310 
\\
                                                                                         & 0.600& 0.600& 0.400& 0.520&  322,197 &  1,499 &  248 &  441 &  19,281 &  248 
\\
                                                                                         & 0.800& 0.800& 0.400& 0.520&  324,451 &  753 &  124 &  459 &  17,992 &  124 
\\
                                                                                         & 0.900& 0.900& 0.400& 0.520&  325,584 &  378 &  62 &  468 &  17,344 &  62 
\\
                                                                                         & 0.950& 0.950& 0.400& 0.520&  326,153 &  189 &  31 &  472 &  17,018 &  31 
\\ \hline
Vary Ballot Trax/Scout implementation M3          & 0.010& 0.900& 0.010& 0.520&  315,984 &  3,250 &  614 &  422 &  23,205 &  614 
\\
with in-person absentee availability M6                                                                                         & 0.027& 0.900& 0.027& 0.520&  316,146 &  3,214 &  604 &  422 &  23,092 &  604 
\\
                                                                                         & 0.050& 0.900& 0.050& 0.520&  316,379 &  3,162 &  590 &  422 &  22,931 &  590 
\\
                                                                                         & 0.400& 0.900& 0.400& 0.520&  319,963 &  2,236 &  372 &  423 &  20,558 &  372 
\\
                                                                                         & 0.500& 0.900& 0.500& 0.520&  321,030 &  1,921 &  310 &  424 &  19,889 &  310 
\\
                                                                                         & 0.600& 0.900& 0.600& 0.520&  322,115 &  1,582 &  248 &  426 &  19,222 &  248 
\\
                                                                                         & 0.800& 0.900& 0.800& 0.520&  324,343 &  837 &  124 &  430 &  17,900 &  124 
\\
                                                                                         & 0.900& 0.900& 0.900& 0.520&  325,485 &  430 &  62 &  433 &  17,243 &  62 
\\
                                                                                         & 0.950& 0.900& 0.950& 0.520&  326,063 &  218 &  31 &  434 &  16,916 &  31 \\ \hline
\end{tabular}%
}
\end{table*}